\documentclass[aps,pra,twocolumn,superscriptaddress]{revtex4}

\usepackage{amsmath}
\usepackage{amssymb}
\usepackage{psfrag}
\usepackage{graphicx}
\usepackage{dcolumn}
\usepackage{bm}
\usepackage{epsfig}
\usepackage{yfonts}
\usepackage{version}
\usepackage[utf8]{inputenc}

\usepackage{color}

\newcommand{\bra}[1]{\langle\,{#1}\, |}
\newcommand{\ket}[1]{|\,{#1}\,\rangle}
\newcommand{\braket}[2]{\mbox{$\langle\,{#1}\, | \,{#2}\,\rangle$}}

\newcommand{\Imag}{\mbox{Im}}
\newcommand{\Real}{\mbox{Re}}

\newcommand{\tH}{\tilde{H}}
\newcommand{\tomega}{\tilde{\omega}}
\newcommand{\ta}{\tilde{a}}
\newcommand{\tkappa}{\tilde{\kappa}}
\newcommand{\talpha}{\tilde{\alpha}}
\newcommand{\tM}{\tilde{M}}
\newcommand{\tpsi}{\tilde{\psi}}

\newcommand{\tA}{\tilde{A}}
\newcommand{\tNull}{\tilde{0}}
\newcommand{\tPsi}{\tilde{\Psi}}
\newcommand{\tPhi}{\tilde{\Phi}}
\newcommand{\tz}{\tilde{z}}
\newcommand{\tL}{\tilde{L}}

\newcommand{\hNull}{\hat{0}}
\newcommand{\homega}{\hat{\omega}}
\newcommand{\hz}{\hat{z}}
\newcommand{\hxi}{\hat{\xi}}
\newcommand{\hO}{\hat{O}}

\newcommand{\sig}{\sigma}

\newlength{\mylenunit}
\begin{document}

\title{Non-Markovian quantum state diffusion for absorption spectra of molecular aggregates}

\author{Jan Roden}
\affiliation{%
Max-Planck-Institut f\"{u}r
Physik komplexer Systeme,
N\"othnitzer Str.\ 38,
D-01187 Dresden, Germany
}%

\author{Walter T.\ Strunz}
\affiliation{Institut f\"{u}r Theoretische Physik,
Technische Universit\"at Dresden, D-01062 Dresden, Germany}

\author{Alexander Eisfeld}
\email{eisfeld@mpipks-dresden.mpg.de}
\affiliation{%
Max-Planck-Institut f\"{u}r
Physik komplexer Systeme,
N\"othnitzer Str.\ 38,
D-01187 Dresden, Germany
}%

\date{\today}

\begin{abstract}
In many molecular systems one encounters the situation where electronic excitations couple to a quasi-continuum of phonon modes.
That continuum may be highly structured e.g.\ due to some weakly damped high frequency modes.
To handle such a situation, an approach combining the non-Markovian quantum state diffusion (NMQSD) description of open quantum systems with an efficient but abstract approximation was recently applied to calculate energy transfer and absorption spectra of molecular aggregates~[Roden, Eisfeld, Wolff, Strunz, PRL 103 (2009) 058301].
To explore the validity of the used approximation for such complicated systems, in the present work we compare the calculated (approximative) absorption spectra with exact results. 
These are obtained from the method of pseudomodes, which we show to be capable of determining the exact spectra for small aggregates and a few pseudomodes.  
It turns out that in the cases considered, the results of the two approaches
mostly agree quite well.
The advantages and disadvantages of the two approaches are discussed.
\end{abstract}


\maketitle

\section{Introduction}

There is growing interest in systems composed of individual monomers that interact via resonant dipole-dipole interaction.
Upon electronic excitation this transition dipole-dipole interaction between the monomers
is responsible for a collective behaviour of these systems.
Besides the classical examples like Van-der-Waals crystals~\cite{Da62__,ScWo06__}, aggregates of organic dyes~\cite{Sc36_563_,Je36_1009_,Sc38_1_,FrTe38_861_,Ko96__}, and light harvesting units
of plants, algae, and bacteria~\cite{FrTe38_861_,AmVaGr00__,ReMaKue01_137_,GrNo06_793_,FrRaeTi09_102_} many new systems have emerged. 
Examples are ultra-cold Rydberg atoms~\cite{RoHeTo04_042703_,AnVeGa98_249_,MueBlAm07_090601_,AtEiRo08_045030_,WAtEi__}, quantum dots~\cite{MaLeXu04_318_,LoReNa03_136_}, assemblies of nano-particles~\cite{ReHoLe03_137_,MaMaKn08_2369_}, and recently also hybrid systems~\cite{HaTiNa09_9986_}. 

The common approach to describe these aggregates is to treat the monomers as
electronic two-level systems. 
Besides the electronic degrees of freedom, however, one often has to take into account nuclear degrees of freedom explicitly.
For instance in the case of molecular aggregates~\cite{Ko96__,EiKnKi09_658_,EiBr06_113003_,WeSt05_1171_,PrDiNi04_097403_,SeLoWue07_6214_}, which will serve as the primary
example in this work, the electronic excitation of a monomer couples strongly
to internal vibrational modes of the monomer and to modes of the surroundings.

Often, the monomer spectrum is dominated by one vibrational progression which is considerably broadened. 
A commonly applied approximation is then to only consider one effective mode
corresponding to this progression~\cite{FuGo61_1059_,KoHaKa81_498_} and to take the broadening into account by convoluting with some lineshape function which is usually assumed to be Gaussian. 
It has been shown that using this approach already important features of experimental spectra can be reproduced~\cite{KoHaKa81_498_,EiBrSt05_134103_,SeMaEn06_354_,EiBr06_376_,Sc95_451_,Sp00_7_}.
Although the resulting spectra reveal many characteristics of the aggregates, important aspects like the detailed shapes of the J-band~\cite{EiBr07_354_} and the H-band~\cite{RoEiBr08_258_} cannot adequately be described by considering only one vibrational mode. 
On the other hand, the exact inclusion of only {\it one} vibrational mode already
complicates the treatment of molecular aggregates considerably, so that this
approach is restricted to small aggregates.
This problem becomes even more serious when one attempts to include more modes in
this manner. 

When the interaction of the  chromophore monomers with the environment
is negligible (e.g.\ in high resolution spectroscopy in helium
nanodroplets~\cite{WeSt05_1171_,RoEiDv10_submitted}), then the {\it explicit} inclusion of vibrational modes is of
great importance.
However, for typical spectra in solution
or in a solid state matrix (where a strong coupling between the chromophores
and the environment is present) it seems better to use a continuum of
vibrations that couple to the electronic excitation to account for the large number of environmental degrees of freedom.

This interaction between the electronic excitation and the vibrations is conveniently
encoded in the so-called spectral density.
It describes the frequency-dependent coupling between the system (the electronic degrees of freedom) and the (continuum of) harmonic oscillators.
In the Markov case the spectral density is assumed to be flat in the relevant frequency regions.
Clearly, for the considered monomers this assumption does not hold. 
Due to strong interaction with some internal vibrational modes, the spectral density will
be highly structured (i.e.\ frequency-dependent), indicating that a non-Markovian theoretical framework is required.

An approach to tackle this complicated problem was recently presented in Ref~\cite{RoEiWo09_058301_}.
The method is based on the non-Markovian quantum state diffusion (NMQSD) description of open quantum systems~\cite{DiSt97_569_}.
Here, the system part is chosen to contain only the electronic degrees of
freedom which interact with a non-Markovian
environment (the bath) comprising all vibrations (internal modes of the monomers as well as external modes).
One then can derive a stochastic evolution equation for states in the (small) space of the system part.
However, solving the exact evolution equation turns out to be very difficult
due to the appearance of a functional derivative w.r.t.\ functionals containing the bath degrees of freedom.
To overcome these difficulties, in an approximation only the zeroth order of a functional expansion (we will refer to it as ZOFE approximation) of the problematic term is taken into account~\cite{YuDiGi99_91_,RoEiWo09_058301_}.
For several (simple) problems this procedure has been shown to give the exact result~\cite{DiGiSt98_1699_,StDiGi99_1801_}.
However, for more complex problems like the molecular aggregates studied in this work, the range of validity of the approximation is not clear.
It should be noted that the NMQSD approach in combination with the ZOFE approximation provides a very efficient calculation scheme: 
in order to obtain the absorption spectrum of the aggregate and the energy transfer between the monomers, the equations one has to solve are in the small Hilbert space of the electronic degrees of freedom alone~\cite{RoEiWo09_058301_}.

One aim of the present paper is to examine the validity of the ZOFE approximation leading to the calculation scheme presented in Ref.~\cite{RoEiWo09_058301_}. 
To this end, we compare with an approach where so-called
pseudomodes~\cite{Im94_3650_,Ga97_2290_,MaMaPi09_012104_} are included into
the system part together with the electronic degrees of freedom. 
The electronic degrees of freedom now couple only to the pseudomodes, the
pseudomodes in turn are then coupled to a Markovian bath. 
For a spectral density consisting of a sum of Lorentzians the pseudomode
method is exact (taking one pseudomode for each Lorentzian).
This allows to directly compare the approximative NMQSD-ZOFE treatment with exact calculations.
However, due to the inclusion of the pseudomodes into the system part, the numerical solution of the corresponding evolution equation is limited to a rather small number of monomers in the aggregate with only a few pseudomodes, i.e.\ only a few Lorentzians in the spectral density.

Besides the possibility of comparing the NMQSD-ZOFE approach with exact calculations,
the pseudomode method has also some physical significance: one can think of
the pseudomodes as internal vibrational modes of a chromophore that strongly
couple to the electronic excitation and which are damped by the coupling to
the remaining vibrations.

The comparison between zero temperature absorption spectra of small aggregates calculated using
the NMQSD-ZOFE approach and spectra obtained from the exact pseudomode
approach shows that in the cases considered there is mostly quite good agreement between
the two approaches.
We will discuss in which situations the approximative result of the NMQSD-ZOFE approach is expected to deviate from the exact solution.

The structure of this paper is as follows:
In Section~\ref{mod_of_agg}, we introduce the Hamiltonian of the aggregate. 
The Hamiltonian is written as the sum of a system part (containing only electronic degrees of
freedom), an environmental part (containing all vibrational modes), and the part of the
interaction between electronic degrees of freedom and vibrations. 
In the following Section~\ref{sec:AbsOfAgg}, the basic formulas that are used to calculate the absorption spectrum are given by specifying the initial state and introducing the dipole
correlation function.
In Section~\ref{sec:qsd_approach}, the general Non-Markovian Schr\"odinger
Equation (NMQSD) approach is applied to the case of an aggregate.
It is shown how the absorption spectrum can be obtained in this approach.
Next, in Section~\ref{sec_zofe}, the ZOFE approximation is introduced.
Then, in Section~\ref{sec:Exact_solv_model}, the pseudomode (PM) approach is
presented.
In Section~\ref{sec:compar_nmqsd_pm}, the NMQSD-ZOFE absorption spectra  are compared with exact PM spectra.
We conclude in Section~\ref{conclusion} by summarizing our findings.
Details of the calculations and minor results have been placed in the appendices.
In Appendix~\ref{ap_exact_solvable}, two exactly solvable cases (namely that of
non-interacting monomers and the case where the coupling to the vibrations can
be considered to be Markovian) are discussed.
In Appendix \ref{sec:ap_absorpPM}, it is  shown how to obtain the absorption
spectrum using the PM approach. The numerical implementation is discussed. 
Finally, in Appendix \ref{sec:ap_equiv_models_with_and_without_pm}, it is shown that for
a Lorentzian spectral density the absorption spectrum obtained from the exact NMQSD approach 
is equal to the spectrum obtained from the PM method.

\section{The aggregate Hamiltonian}
\label{mod_of_agg}

We consider an aggregate consisting of $N$ monomers, labelled by $n=1,\ldots,N$.
For each monomer $n$ we take into account its electronic ground state $\ket{\phi_n^g}$ 
and one excited electronic state $\ket{\phi_n^e}$.
The transition energy between these two states (whose wave functions we take
to be real) is denoted by $\varepsilon_n$.
Apart from the two electronic states, each monomer has a collection of vibrational modes 
comprising internal modes as well as modes of the local environment of the monomer.
We will refer to these degrees of freedom as ``nuclear coordinates''.
The electronic excitation of monomer $n$ couples to its vibrations and we choose the 
vibrational modes to be harmonic and the coupling to be linear (making contact to 
previous work ~\cite{WiMo60_872_,Me63_154_,FuGo64_2280_,Wi65_161_,ScFi84_269_,EiBrSt05_134103_,SeMaEn06_354_,WaEiBr08_044505_}).
The Hamiltonian of monomer $n$ is then given by
\begin{equation}
  H_n=H^g_n\ket{\phi_n^g}\bra{\phi_n^g}+H^e_n\ket{\phi_n^e}\bra{\phi_n^e},
\end{equation} 
with the Hamiltonian of the vibrations in the electronic ground state
\begin{equation}
\label{HamMonGround}
  H^g_n=\sum_{\lambda}\hbar\omega_{n\lambda}a^{\dagger}_{n\lambda}a_{n\lambda}
\end{equation}
(the energies of the vibrational ground states of all modes in the electronic ground 
state are chosen to be zero). The Hamiltonian
of the vibrations in the excited electronic state comprising a shift,
reads
\begin{equation}
\label{HamMonExc}
  H^e_n=\varepsilon_n+\sum_{\lambda}\hbar\omega_{n\lambda}a^{\dagger}_{n\lambda}a_{n\lambda}-\sum_{\lambda}\kappa_{n\lambda}(a^{\dagger}_{n\lambda}+a_{n\lambda}).
\end{equation}
Here $a_{n\lambda}$ denotes the annihilation operator of mode $\lambda$ of monomer $n$ with frequency $\omega_{n\lambda}$. 
The coupling strength with which electronic excitation of monomer $n$ couples to mode $\lambda$ of this monomer is denoted by $\kappa_{n\lambda}$.

For the aggregate we assume that the electronic wave functions of the monomers do not overlap.
The electronic ground state of the aggregate is then taken as the product
\begin{equation}
\label{ElGroundStateAgg}
 \ket{g_{\rm el}}=\prod_{m=1}^N \ket{\phi^g_m}
\end{equation}
of the electronic ground states $\ket{\phi_m^g}$ of all monomers.
A state of the aggregate in which only monomer $n$ is electronically excited and all other monomers are in their electronic ground state we denote by
\begin{equation}
\label{pi_n_states}
 \ket{\pi_n}=\ket{\phi^e_n}\prod_{m\ne n}^N \ket{\phi^g_m}.
\end{equation}
We expand the aggregate Hamiltonian w.r.t.\ the states 
Eqs.~(\ref{ElGroundStateAgg}) 
and (\ref{pi_n_states}) and neglect states with more than one electronic excitation 
on the aggregate. 
Thus we obtain the Hamiltonian
\begin{equation}
\label{AggHamExpandGroundAndExcAggState}
  H=H^g\ket{g_{\rm el}}\bra{g_{\rm el}}+H^e\sum_{n=1}^N \ket{\pi_n}\bra{\pi_n}
\end{equation}
for the aggregate, with the part 
\begin{equation}
  H^g=\sum_{n=1}^N H_n^g 
\end{equation}
for the electronic ground state and the part
\begin{equation}
\label{HamTotExc}
\begin{split}
   H^e=\sum_{n=1}^N &\left(H_n^e+\sum_{m\ne n}^N H_m^g\right)\ket{\pi_n}\bra{\pi_n}\\
&+ \sum_{n,m=1}^N V_{nm}\ket{\pi_n}\bra{\pi_m} 
\end{split}
\end{equation}
for the electronically excited state.
The matrix element $V_{nm}$, causing electronic excitation to be transfered
from monomer $n$ to monomer $m$ via transition dipole-dipole interaction, is 
taken to be independent of nuclear coordinates (note that $H_n^g$ and $H_n^e$ depend on 
nuclear coordinates through Eqs.~(\ref{HamMonGround}) and (\ref{HamMonExc})).
With the Hamiltonians of the monomers Eqs.~(\ref{HamMonGround}) and (\ref{HamMonExc}) we can write Eq.~(\ref{HamTotExc}) as
\begin{equation}
\label{HeAsSysPlusIntPlusEnv}
  H^e=H_{\rm sys}+H_{\rm int}+H_{\rm env},
\end{equation}
with the purely electronic ``system'' part
\begin{equation}
\label{H_sys}
  H_{\rm sys}=\sum_{n=1}^N\varepsilon_n\ket{\pi_n}\bra{\pi_n}+\sum_{n,m=1}^N V_{nm}\ket{\pi_n}\bra{\pi_m},
\end{equation}
and a part $H_{\rm env}$ describing the ``environment'' of vibrational modes
\begin{equation}
\label{H_env}
  H_{\rm env}=\sum_{n=1}^N\sum_{\lambda}\hbar\omega_{n\lambda}a^{\dagger}_{n\lambda}a_{n\lambda}.
\end{equation}
 The coupling of electronic excitation to these vibrations is expressed through
\begin{equation}
\label{HInt}
  H_{\rm int}=-\sum_{n=1}^N\ket{\pi_n}\bra{\pi_n}\sum_{\lambda}\kappa_{n\lambda}(a^{\dagger}_{n\lambda}+a_{n\lambda}).
\end{equation}
We emphasize that with Eqs.~(\ref{H_sys})-(\ref{HInt}) we make a special choice of the three parts, 
system, interaction and environment, 
of the aggregate Hamiltonian.
In particular the system part Eq.~(\ref{H_sys}) contains only electronic degrees of freedom.
In Section~\ref{sec:Exact_solv_model}, where we will introduce the pseudomode approach, we will  
include also vibrational modes (pseudomodes) into the system part. 

A useful quantity describing many aspects of the coupling of the system degrees of 
freedom to the vibrational environment is the so-called spectral density
\cite{MaKue00__}, which for monomer 
$n$ is given by
\begin{equation}
\label{spec_dens}
J_n(\omega)=\sum_{\lambda}|\kappa_{n\lambda}|^2\ \delta(\omega-\omega_{n\lambda})
\end{equation}
and which will be used later, generalized to a continuum of vibrational modes.


\section{Absorption of the aggregate}
\label{sec:AbsOfAgg}

We consider absorption of light by the aggregate at zero temperature.
Initially, the aggregate is taken to be in its total ground state
\begin{equation}
\label{InitAggStateAbs}
  \ket{\Phi(t=0)}=\ket{g_{\rm el}}\ket{0},
\end{equation}
which is a product of the electronic ground state $\ket{g_{\rm el}}$ defined in 
Eq.~(\ref{ElGroundStateAgg}) and the ground state $\ket{0}=\prod_{n\lambda}\ket{0_{n\lambda}}$ of $H_{\rm env}$, 
i.e.\ all vibrational modes of all monomers are in their ground state
$\ket{0_{n\lambda}}$.
The absorption of light with frequency $\nu$ and polarization 
$\vec{\mathcal{E}}$ is then given by \cite{MaKue00__}
\begin{equation}
\label{AbsSpec}
  A(\nu)= \mbox{Re}\int_0^{\infty}dt\ e^{i\nu t}M(t).
\end{equation}
Here,
\begin{equation}
\label{GeneralTimeCorrFct}
  M(t)=\bra{\Phi(t=0)}\hat{\vec{\mu}}\cdot\vec{\mathcal{E}}\ e^{-iHt/\hbar}\ \hat{\vec{\mu}}\cdot\vec{\mathcal{E}}^*\ket{\Phi(t=0)}
\end{equation} 
is the dipole correlation function
where $\hat{\vec{\mu}}$ denotes the total dipole operator of the aggregate, given by the sum
\begin{equation}
\label{TotalDipSumOfMonDip}
  \hat{\vec{\mu}}=\sum_{n=1}^N \hat{\vec{\mu}}_n
\end{equation}
of all monomer dipole operators $\hat{\vec{\mu}}_n$. Note that
the Hamiltonian $H$ in the exponent of the propagator in Eq.~(\ref{GeneralTimeCorrFct}) is the aggregate Hamiltonian of Eq.~(\ref{AggHamExpandGroundAndExcAggState}).
We take $\hat{\vec{\mu}}_n$ to be independent of nuclear coordinates.
Then the correlation function Eq.~(\ref{GeneralTimeCorrFct}) with Eq.~(\ref{InitAggStateAbs}) reads
\begin{equation}
\label{DipCorrFuncPureElDipoles}
  M(t)=\sum_{n,m=1}^N(\vec{\mu}_m^*\cdot\vec{\mathcal{E}})\bra{\pi_m}\bra{0}\
  e^{-iHt/\hbar}\ \ket{0} \ket{\pi_n}(\vec{\mu}_n\cdot\vec{\mathcal{E}}^*),
\end{equation}
with the monomer transition dipoles 
\begin{equation}
\label{MonTransDip}
  \vec{\mu}_n\equiv\bra{\phi_n^e}\hat{\vec{\mu}}_n\ket{\phi_n^g}.
\end{equation}
Defining
\begin{equation}
\label{DefMuPrefac}
  \mu_{\rm tot}\equiv\sqrt{\sum_{n=1}^N|\vec{\mu}_n\cdot\vec{\mathcal{E}}^*|^2},
\end{equation}
equation~(\ref{DipCorrFuncPureElDipoles}) can be written as
\begin{equation}
\label{MofTwithMuPrefactor}
  M(t)=\mu_{\rm tot}^2\braket{\Psi(t=0)}{\Psi(t)}
\end{equation}
with
\begin{equation}
\label{PsiOftPropagator}
  \ket{\Psi(t)}=\exp(-iH^e t/\hbar)\ket{\Psi(t=0)},
\end{equation}
where $H^e$ is given by Eq.~(\ref{HamTotExc})
and the initial state is
\begin{equation}
\label{DefPsiTimeZero}
  \ket{\Psi(t=0)}=\ket{\psi_0}\ket{0}.
\end{equation}
The normalized initial electronic state $\ket{\psi_0}$ in Eq.~(\ref{DefPsiTimeZero}) is given by
\begin{equation}
\label{DefPsiNull}
  \ket{\psi_0}= \frac{1}{\mu_{\rm tot}}\sum_{n=1}^N(\vec{\mu}_n\cdot\vec{\mathcal{E}}^*)\ket{\pi_n}
\end{equation}
and contains explicitly the geometry of the aggregate via the orientation of the transition dipoles $\vec{\mu}_n$ of the monomers.
In the following, our goal will be to obtain the state $\ket{\Psi(t)}$ to be able to calculate the absorption spectrum according to Eqs.~(\ref{AbsSpec}) and (\ref{MofTwithMuPrefactor}).


\section{The general NMQSD approach}
\label{sec:qsd_approach}

The correlation function $M(t)$ of Eq.~(\ref{MofTwithMuPrefactor})
can be obtained using the framework of stochastic Schr\"odinger equations,
here the non-Markovian quantum state diffusion (NMQSD) approach (Ref.~\cite{DiSt97_569_}).
Note, however, that no stochasticity will enter in the following calculations due
to the fact that for a zero temperature absorption spectrum considered in this work,
one has to project on the environmental ground state (see section \ref{sec:abs_in_qsd}). 
For energy transfer between monomers, however, the full stochasticity of the
non-Markovian quantum state diffusion (NMQSD) approach (Ref.~\cite{DiSt97_569_})
resurfaces \cite{RoEiWo09_058301_}.

\subsection{NMQSD evolution equation for system states}

We briefly summarize the NMQSD approach as applied to molecular aggregates.
First, we transform $H^e$ to the interaction representation w.r.t. $H_{\rm env}$, yielding
\begin{equation}
\label{H_e_Interac_repres}
  H^e(t)=H_{\rm sys}+\sum_{n=1}^N\left(L_nA^{\dagger}_n(t)+L^{\dagger}_nA_n(t)\right)
\end{equation}
with
\begin{equation}
  L_n\equiv -\ket{\pi_n}\bra{\pi_n}=L^{\dagger}_n
\end{equation}
and
\begin{equation}
  A_n(t)\equiv\sum_{\lambda}\kappa_{n\lambda}a_{n\lambda}e^{-i\omega_{n\lambda}t}.
\end{equation}
Thus, we have the time-dependent Schr\"odinger equation
\begin{equation}
\label{SchroeEqPsiTotal}
  i\hbar\partial_t\ket{\Psi(t)}=H^e(t)\ket{\Psi(t)}
\end{equation}
in the interaction picture.
In the next step we expand the total wave function $\ket{\Psi(t)}$ w.r.t.\ Bargmann coherent states $\ket{z_{n\lambda}}=\exp(z_{n\lambda}a^{\dagger}_{n\lambda})\ket{0_{n\lambda}}$~\cite{Ba61_187_} of the vibrations, where $\ket{0_{n\lambda}}$ is the ground state of vibrational mode $\lambda$ of monomer $n$ and the $z_{n\lambda}$ are complex numbers.
One then obtains
\begin{equation}
\label{ExpandTotalPsiWithBargmann}
  \ket{\Psi(t)}=\int\frac{d^2{\mathbf z}}{\pi}e^{-|{\mathbf z}|^2}\ket{\psi(t,{\mathbf z}^*)}\ket{{\mathbf z}},
\end{equation}
with states $\ket{\psi(t,{\mathbf z}^*)}$ in the space of the electronic system $H_{\rm sys}$, $\ket{{\mathbf z}}=\prod_n\prod_{\lambda}\ket{z_{n\lambda}}$, and $d^2z=d\, \Real(z)\ d\,\Imag( z)$.
Inserting Eq.~(\ref{ExpandTotalPsiWithBargmann}) into Eq.~(\ref{SchroeEqPsiTotal}) one finds \cite{DiSt97_569_} that the states $\ket{\psi(t,{\mathbf z}^*)}$ appearing in Eq.~(\ref{ExpandTotalPsiWithBargmann}) obey an evolution equation
\begin{equation}
\label{EvolEqPsiSysSpace}
\begin{split}
\partial_t \ket{\psi(t,{\mathbf z}^*)}&=
-\frac{i}{\hbar} H_{\rm sys}\ket{\psi(t,{\mathbf z}^*)}
 +\sum_n L_n z^*_{t,n}\ket{\psi(t,{\mathbf z}^*)}\\
& -\frac{1}{\hbar^2}\sum_n L_n^{\dagger} \int_0^t ds\ \alpha_n(t-s) \frac{\delta \ket{\psi(t,{\mathbf z}^*)} }{\delta z^*_{s,n} }
\end{split}
\end{equation} 
in the small Hilbert space of the electronic degrees of freedom alone,
with time-dependent complex numbers
\begin{equation}
  z^*_{t,n}=-\frac{i}{\hbar}\sum_{\lambda}\kappa_{n\lambda}z^*_{n\lambda}e^{i\omega_{n\lambda}t}
\end{equation}
and where 
\begin{equation}
\label{bathCorrFuncTZero}
\begin{split}
  \alpha_n(t-s)&=\left\langle A_n(t)A_n^{\dagger}(s)\right\rangle_{T=0}= \bra{0}A_n(t)A_n^{\dagger}(s)\ket{0}\\
  &=\sum_{\lambda}|\kappa_{n\lambda}|^2e^{-i\omega_{n\lambda}(t-s)}
\end{split}
\end{equation}
is the zero temperature bath correlation function of monomer $n$, which encodes the interaction of an electronic excitation with the environment of vibrations.
Note that in this case of zero temperature, $\alpha_n(t-s)$ is just the Fourier transform of the 
spectral density Eq.~(\ref{spec_dens}), i.e.\ $\alpha_n(\tau)=\int d\omega\  e^{-i \omega\tau}J_n(\omega)$.

For a non-Markovian bath, the solution
of Eq.~(\ref{EvolEqPsiSysSpace}) is complicated due to the appearance of
the memory integral over $\alpha_n(t-s)$ and the appearance of a functional derivative as
an integrand.
To deal with the functional derivative we follow Ref.~\cite{DiGiSt98_1699_} and write
\begin{equation}
\label{DefOfO}
  \frac{\delta}{\delta z^*_{s,n} } \ket{\psi(t,{\mathbf z}^*)}=O^{(n)}(t,s,{\mathbf z}^*)\ket{\psi(t,{\mathbf z}^*)}
\end{equation}
with an operator $O^{(n)}$ acting in the space of the electronic system part $H_{\rm sys}$ only.
The operators $O^{(n)}(t,s,{\mathbf z}^*)$ have to obey the consistency condition
\begin{equation}
\label{consis_condition}
  \partial_t\left(O^{(n)}(t,s,{\mathbf z}^*)\ket{\psi(t,{\mathbf z}^*)}\right)=\frac{\delta}{\delta z^*_{s,n} }\,\partial_t\ket{\psi(t,{\mathbf z}^*)},
\end{equation}
which leads, using Eq.~(\ref{EvolEqPsiSysSpace}) and introducing the abbreviation
\begin{equation}
\label{DefOQuer}
  \bar{O}^{(n)}(t,{\mathbf z}^*)=\frac{1}{\hbar^2}\int_0^t ds\ \alpha_n(t-s)O^{(n)}(t,s,{\mathbf z}^*),
\end{equation} 
to an evolution equation~\cite{StYu04_052115_} 
\begin{equation}
\label{EvolEqOOpWithFuncDer}
\begin{split}
\partial_t O^{(m)}&(t,s,{\mathbf z}^*)=\left[-\frac{i}{\hbar}H_{\rm sys},O^{(m)}(t,s,{\mathbf z}^*)\right]\\
&+\sum_n\left[L_nz^*_{t,n}-L^{\dagger}_n\bar{O}^{(n)}(t,{\mathbf z}^*),O^{(m)}(t,s,{\mathbf z}^*)\right]\\
&-\sum_nL^{\dagger}_n\frac{\delta \bar{O}^{(n)}(t,{\mathbf z}^*)}{\delta z^*_{s,m}}.
\end{split}
\end{equation}
The latter has to be solved with initial condition
\begin{equation}
\label{init_cond_OOp}
  O^{(n)}(s,s,{\mathbf z}^*)=L_n.
\end{equation}
Finally, using Eq.~(\ref{DefOfO}) and (\ref{DefOQuer}), the evolution equation (\ref{EvolEqPsiSysSpace}) 
turns into the linear non-Markovian QSD equation
\begin{equation}
\label{EvolEqPsiSysSpaceMitOQuer}
\begin{split}
  \partial_t&\ket{\psi(t,{\mathbf z}^*)}=-\frac{i}{\hbar} H_{\rm sys}\ket{\psi(t,{\mathbf z}^*)}\\
&+\sum_n\left(L_nz^*_{t,n}-L^{\dagger}_n\bar{O}^{(n)}(t,{\mathbf z}^*)\right)\ket{\psi(t,{\mathbf z}^*)}.
\end{split}
\end{equation}
Due to the functional derivative appearing in Eq.~(\ref{EvolEqOOpWithFuncDer}),
the operator $\bar{O}^{(n)}(t,{\mathbf z}^*)$ Eq.~(\ref{DefOQuer}) cannot be evaluated in the general case.
However, it can be obtained in some special cases, e.g.\ in  the Markovian case (see Appendix~\ref{ap_markovian_bath}) and the case of non-interacting monomers (see Appendix~\ref{ap_non_interac_monomers}).

\subsection{Absorption in the NMQSD approach}
\label{sec:abs_in_qsd}

Using the NMQSD approach, the correlation function 
$M(t)$ of Eq.~(\ref{MofTwithMuPrefactor}) can be calculated as follows.
Inserting the expansion Eq.~(\ref{ExpandTotalPsiWithBargmann}) into Eq.~(\ref{MofTwithMuPrefactor}) yields
\begin{equation}
\label{MofTwithIntBargmann}
  M(t)=\mu_{\rm tot}^2\int\frac{d^2{\mathbf z}}{\pi}e^{-|{\mathbf z}|^2}\braket{\psi_0}{\psi(t,{\mathbf z}^*)}\braket{0}{{\mathbf z}},
\end{equation}
where $\ket{\psi_0}$ is given by Eq.~(\ref{DefPsiNull}). Here we used that 
$\bra{0}e^{-i H_{\rm env} t/\hbar}$ -- which appears through the transformation to the interaction 
representation -- is equal to $\bra{0}$. From
equation~(\ref{MofTwithIntBargmann}) we get~\cite{Ba61_187_}
\begin{equation}
\label{M_of_t_QSD}
  M(t)=\mu_{\rm tot}^2\braket{\psi_0}{\psi(t,{\mathbf z}^*=0)}
\end{equation}
where $\ket{\psi(t,{\mathbf z}^*=0)}$ can be obtained using Eq.~(\ref{EvolEqPsiSysSpaceMitOQuer}) with ${\mathbf z}^*=0$, i.e.\ 
\begin{equation}
\label{EvolEqPsiSysSpaceMitOQuerZeqZero}
\begin{split}
  \partial_t&\ket{\psi(t,{\mathbf z}^*=0)}=\\
&\left(-\frac{i}{\hbar}H_{\rm sys}-\sum_nL^{\dagger}_n\bar{O}^{(n)}(t,{\mathbf z}^*=0)\right)\ket{\psi(t,{\mathbf z}^*=0)}
\end{split}
\end{equation}
with the initial condition $\ket{\psi(t=0,{\mathbf z}^*=0)}=\ket{\psi_0}$.
Note that due to the appearance of the functional derivative in Eq.~(\ref{EvolEqOOpWithFuncDer}) the determination of $\bar{O}^{(n)}(t)$ is still a formidable task.
 
\section{The zeroth order functional expansion (ZOFE) approximation}
\label{sec_zofe}

The main task in the NMQSD approach is to obtain the operator $\bar{O}^{(n)}(t,{\mathbf z}^*)$ 
(and $O^{(n)}(t,s,{\mathbf z}^*)$ respectively) being defined in Eq.~(\ref{DefOQuer}) 
(and Eq.~(\ref{DefOfO})).
An evolution equation for $O^{(n)}(t,s,{\mathbf z}^*)$ is given by Eq.~(\ref{EvolEqOOpWithFuncDer}).
Although for the calculation of the zero temperature absorption spectrum from 
Eq.~(\ref{EvolEqPsiSysSpaceMitOQuerZeqZero}) only the values of $O^{(n)}(t,s,{\mathbf z}^*)$ 
for ${\mathbf z}^*=0$ are needed, Eq.~(\ref{EvolEqOOpWithFuncDer}) contains the 
${\mathbf z}^*$ dependence via the functional derivative in a non-local way.
To simplify Eq.~(\ref{EvolEqOOpWithFuncDer}) we follow Ref.~\cite{YuDiGi99_91_} expanding the 
operator $O^{(n)}(t,s,{\mathbf z}^*)$ w.r.t. $z^*_{t,n}$ in a functional way and 
keep only the zeroth order term of the functional expansion. In other words, we approximate
\begin{equation}
\label{NoNoiseApprox}
  O^{(n)}(t,s,{\mathbf z}^*)\approx O^{(n)}_0(t,s)
\end{equation}
to be independent of ${\mathbf z}^*$ and refer to Eq.~(\ref{NoNoiseApprox})
as the zeroth order functional expansion (ZOFE) approximation. 
Then, from Eq.~(\ref{EvolEqOOpWithFuncDer}) one obtains an approximate evolution equation~\cite{YuDiGi99_91_}
\begin{equation}
\begin{split}
  \partial_tO_0^{(n)}(t,s)=&\left[-\frac{i}{\hbar}H_{\rm sys},O^{(n)}_0(t,s)\right]\\
&-\sum_m\left[L_m^{\dagger}\bar{O}^{(m)}_0(t),O^{(n)}_0(t,s)\right]
\end{split}
\end{equation}
with initial condition $O^{(n)}_0(s,s)=L_n$ (where we obtain $\bar{O}^{(n)}_0(t)$ from $O^{(n)}_0(t,s)$ via Eq.~(\ref{DefOQuer})).
Inserting the approximate operator $\bar{O}^{(n)}_0(t)$ 
into Eq.~(\ref{EvolEqPsiSysSpaceMitOQuerZeqZero}) gives
\begin{equation}
\label{EvolEqPsiSysSpaceMitOQuerZeqZeroNNA}
\begin{split}
  \partial_t&\ket{\psi(t,{\mathbf z}^*=0)}=\\
&\left(-\frac{i}{\hbar}H_{\rm sys}-\sum_nL^{\dagger}_n\bar{O}^{(n)}_0(t)\right)\ket{\psi(t,{\mathbf z}^*=0)}
\end{split}
\end{equation}
whose numerical implementation is straightforward~\footnote{for the propagation we use a forth order Runge Kutta method.}.
To obtain the absorption spectrum of the aggregate, we solve equation~(\ref{EvolEqPsiSysSpaceMitOQuerZeqZeroNNA}) for the initial condition $\ket{\psi(0,{\mathbf z}^*=0)}=\ket{\psi_0}$ given by Eq.~(\ref{DefPsiNull}). 
Then, we calculate the spectrum via Eqs.~(\ref{M_of_t_QSD}) and~(\ref{AbsSpec}). 

\section{Pseudomode method}
\label{sec:Exact_solv_model}

As we will show in subsequent sections, the NMQSD approach in combination with the ZOFE
approximation presented in Section~\ref{sec:qsd_approach} 
offers a highly efficient method to 
calculate absorption spectra of molecular aggregates.
However, it relies on the rather abstract ZOFE approximation made in Eq.~(\ref{NoNoiseApprox}).
To obtain information about the accuracy of the approximation, in the following we will 
compare with exact calculations.
It is clear that in general, an exact determination of aggregate spectra is a
formidable task, impossible for arbitrary spectral densities and an arbitrary 
number $N$ of monomers.
In this section we review a method~\cite{Im94_3650_,Ga97_2290_,MaMaPi09_012104_} that allows a numerically exact calculation of spectra 
for small aggregates for bath correlation functions of the form
\begin{equation}
\label{BathCorrFuncForLorentzSpecDens}
  \alpha_n(t-s)=\sum_j\Gamma_{nj}e^{-i\Omega_{nj}(t-s)-\gamma_{nj}|t-s|}
\end{equation}
which corresponds to the spectral density 
\begin{equation}
\label{LorentzSpecDens}
  J_n(\omega)=\frac{1}{\pi}\sum_j\Gamma_{nj}\frac{\gamma_{nj}}{(\omega-\Omega_{nj})^2+\gamma_{nj}^2}
\end{equation}
being a sum of Lorentzians centered at $\Omega_{nj}$ with width $\gamma_{nj}$.
For the numerical implementation of the method, the number of Lorentzians which can be taken into
account is limited (see examples in section \ref{sec:compar_nmqsd_pm}).
As a special case, the limit $\gamma_{nj}\rightarrow 0$ is included, i.e.\ the case of
  undamped vibrational modes, which has been extensively 
  studied in the literature
  \cite{FuGo64_2280_,ScFi84_269_,BoTrBa99_1633_,Sp09_4267_,GuZuCh08_2094_,RoEiBr08_258_,WaEiBr08_044505_}.
From the point of view of open system dynamics, the memory time of the environment is
clearly infinite in such a case.

We want to point out that the NMQSD-ZOFE approach is not restricted to the special form~(\ref{BathCorrFuncForLorentzSpecDens}) for the bath correlation function.
However, note also that in principle an arbitrary bath correlation function can be approximated by a
sum of exponentials in the form of Eq.~(\ref{BathCorrFuncForLorentzSpecDens}) as 
discussed in Refs.~\cite{MeTa99_3365_,Im94_3650_,Ga97_2290_}.

\subsection{The pseudomode (PM) Hamiltonian}
In the pseudomode approach (see e.g.~\cite{Im94_3650_,Ga97_2290_,MaMaPi09_012104_}) the system part of 
the aggregate Hamiltonian is enlarged. Apart from the electronic degrees of freedom, auxiliary
vibrational degrees of freedom (pseudomodes) are included in the ``system'', each coupled to a 
Markovian bath in a way specified below.

The Hamiltonian of the aggregate is written as
\begin{equation}
\label{AggHamPM}
  \tH=\tH^g\ket{g_{\rm el}}\bra{g_{\rm el}}+\tH^e\sum_{n=1}^N\ket{\pi_n}\bra{\pi_n}
\end{equation}
where $\tH^g$ is the vibrational Hamiltonian in the electronic ground state.
The relevant Hamiltonian $\tH^e$ of the aggregate in the excited electronic state 
is given by
\begin{equation}
\label{H_e_PM}
  \tH^e=\tH_{\rm sys}+\tH_{\rm int}+\tH_{\rm env}
\end{equation}
with the following choice of system, interaction and environment:
the system part is chosen to be
\begin{equation}
\label{H_SysPM}
\begin{split}
  \tH_{\rm sys}=&H_{\rm sys}+\sum_{n=1}^N\sum_j\hbar\Omega_{nj}b^{\dagger}_{nj}b_{nj}\\
&+\sum_{n=1}^N\sum_j\sqrt{\Gamma_{nj}}\left(L_nb^{\dagger}_{nj}+L_n^{\dagger}b_{nj}  \right).
\end{split}
\end{equation}
Here, as before, the Hamiltonian $H_{\rm sys}$ is that of the purely electronic system 
given by Eq.~(\ref{H_sys}). Additionally, for each monomer we include a set of vibrational 
modes (second term of Eq.~(\ref{H_SysPM})), where mode $j$ of monomer $n$ has a frequency $\Omega_{nj}$ 
(see Eq.~(\ref{BathCorrFuncForLorentzSpecDens})) and annihilation operator $b_{nj}$.
These modes, enumerated with index $j$, are referred to as pseudomodes (PM).
The third term of Eq.~(\ref{H_SysPM}) describes the coupling of electronic excitation on 
monomer $n$ to its PM $j$ with coupling strength $\Gamma_{nj}$ 
(see Eq.~(\ref{BathCorrFuncForLorentzSpecDens})).
Each of the PMs has its own environment whose modes are enumerated with index $\rho$, so that
in obvious notation,
\begin{equation}
\label{H_EnvPM}
  \tH_{\rm env}=\sum_{n=1}^N\sum_j\sum_{\rho}\hbar\tomega_{nj\rho}\ta_{nj\rho}^{\dagger}\ta_{nj\rho}.
\end{equation}
The PMs interact with their environments through
\begin{equation}
\label{H_IntPM}
  \tH_{\rm int}=\sum_{n=1}^N\sum_j\sum_{\rho}\left(\tkappa_{nj\rho}^*\ta_{nj\rho}b^{\dagger}_{nj}+\tkappa_{nj\rho}\ta_{nj\rho}^{\dagger}b_{nj}  \right),
\end{equation}
where $\tkappa_{nj\rho}$ denotes the coupling strength between PM $j$ of monomer $n$ to mode $\rho$ 
of its local environment.

We now take the bath correlation functions of the PMs to read
\begin{equation}
\label{MarkovBathCorrFuncPM}
  \talpha_{nj}(t-s)=\sum_{\rho}|\tkappa_{nj\rho}|^2 e^{-i\tomega_{nj\rho}(t-s)}\equiv 2\hbar^2\gamma_{nj}\delta(t-s),
\end{equation}
i.e.\ the PMs couple to a Markovian environment 
(the parameter $\gamma_{nj}$ is that of Eq.~(\ref{BathCorrFuncForLorentzSpecDens})).
For a Markovian environment, however, we are in the regime of the standard Markov QSD
of Section~\ref{sec:qsd_approach} and find an exact solution (i.e.\ without applying the ZOFE approximation) 
for the time-dependent aggregate state and the absorption spectrum. Clearly, the price
to pay is the need to propagate in the much larger Hilbert space of electronic and PM degrees
of freedom.

We show in Appendix~\ref{sec:ap_equiv_models_with_and_without_pm} that for a bath 
correlation function of the type of Eq.~(\ref{BathCorrFuncForLorentzSpecDens}), the exact
NMQSD approach without ZOFE approximation of Section~\ref{sec:qsd_approach} and the PM method result in the same 
absorption spectra.

\subsection{Absorption spectrum within the PM approach}
\label{sec:ExactAbsSpec}
As in Eq.~(\ref{AbsSpec}) the absorption spectrum is calculated from
\begin{equation}
\label{AbsSpecPM}
  A(\nu)= \mbox{Re}\int_0^{\infty}dt\ e^{i\nu t}\tM(t),
\end{equation}
where now instead of Eq.~(\ref{M_of_t_QSD}) we have 
\begin{equation}
\label{M_of_t_QSD_PM}
  \tM(t)=\mu_{\rm tot}^2\braket{\tpsi_0}{\tpsi(t)}
\end{equation}
with
\begin{equation}
\label{DefPsiNullPM}
  \ket{\tpsi_0}=\ket{\psi_0}\ket{g_{\rm PM}}.
\end{equation}
Here, the initial electronic state $\ket{\psi_0}$ contains the action of the dipole operator, see
Eq.~(\ref{DefPsiNull}) and $\ket{g_{\rm PM}}$ is a product of the ground states of all PMs.
As shown in detail in Appendix~\ref{sec:ap_absorpPM}, the state $\ket{\tpsi(t)}$ in 
Eq.~(\ref{M_of_t_QSD_PM}) is obtained by solving the Schr\"odinger equation
\begin{equation}
\label{EvolEqToCalcPMAbsSpec}
  \partial_t\ket{\tpsi(t)}=\left( -\frac{i}{\hbar}\tH_{\rm sys}-\sum_{n=1}^N\sum_j\gamma_{nj}b^{\dagger}_{nj}b_{nj} \right)\ket{\tpsi(t)}
\end{equation}
with the initial state $\ket{\tpsi_0}$ and $\tH_{\rm sys}$ from Eq.~(\ref{H_SysPM}).
We describe in Appendix~\ref{sec:ap_absorpPM} how this differential equation is solved numerically.
There, we see that its solution becomes quite involved for large aggregates due to the
explicit inclusion of the PMs into the ``system''.

We note that Eq.~(\ref{EvolEqToCalcPMAbsSpec}) is just the Markov QSD equation with zero noise,
and for the PM method takes the role of
Eq.~(\ref{EvolEqPsiSysSpaceMitOQuerZeqZero}) in the NMQSD approach.

\section{Comparison between ZOFE and  exact PM spectra}
\label{sec:compar_nmqsd_pm}

In this section, we will compare absorption spectra calculated using the ZOFE approximation with numerically exact calculations obtained in the PM approach for spectral densities of the form Eq.~(\ref{LorentzSpecDens}).
While being exact in the PM approach, the price one has to pay here is the increase of the number of degrees of freedom of the system part due to the inclusion of the PM into the system.
This entails a rapid growth of the Hamiltonian matrix as the number $M$ of PMs or the number $N$ of monomers of the aggregate is increased.
Thus, using the PM approach computer capabilities limit us to absorption spectra of aggregates with roughly $N=2$ and $M\approx 6$ or $N=3$ and $M\approx 5$ etc.
The values of $N$ and $M$ we can handle, depend also on the coupling strength $\Gamma_{nj}$ of the PM to the electronic excitation: the larger the coupling  $\Gamma_{nj}$, the more basis states have to be taken into account.

In the following, we consider a linear arrangement of monomers with identical properties, i.e.\ $\Gamma_{nj}=\Gamma_j$, $\Omega_{nj}=\Omega_j$, $\gamma_{nj}=\gamma_j$ for the $j$th PM of all monomers.
For simplicity, we take all transition dipole moments of the monomers to be equal and consider only nearest neighbour interaction between the monomers, which we denote by $V$.
We follow Simpson and Peterson \cite{SiPe57_588_} and speak of strong/intermediate/weak interaction, if $V$ is larger/similar/smaller than the width of the monomer spectrum.
As a parameter for the strength of the coupling of electronic excitation to the PMs we use the dimensionless Huang-Rhys factor~\cite{MeOs95__}
\begin{equation}
  X_j=\Gamma_j/(\hbar\Omega_j)^2.
\end{equation}
This Huang-Rhys factor together with the ratios of the energies $\hbar\Omega_j$, $\hbar\gamma_j$ and $V$ determines the shape of the absorption spectra.

\begin{figure}[h]
  \includegraphics[width=1\mylenunit]{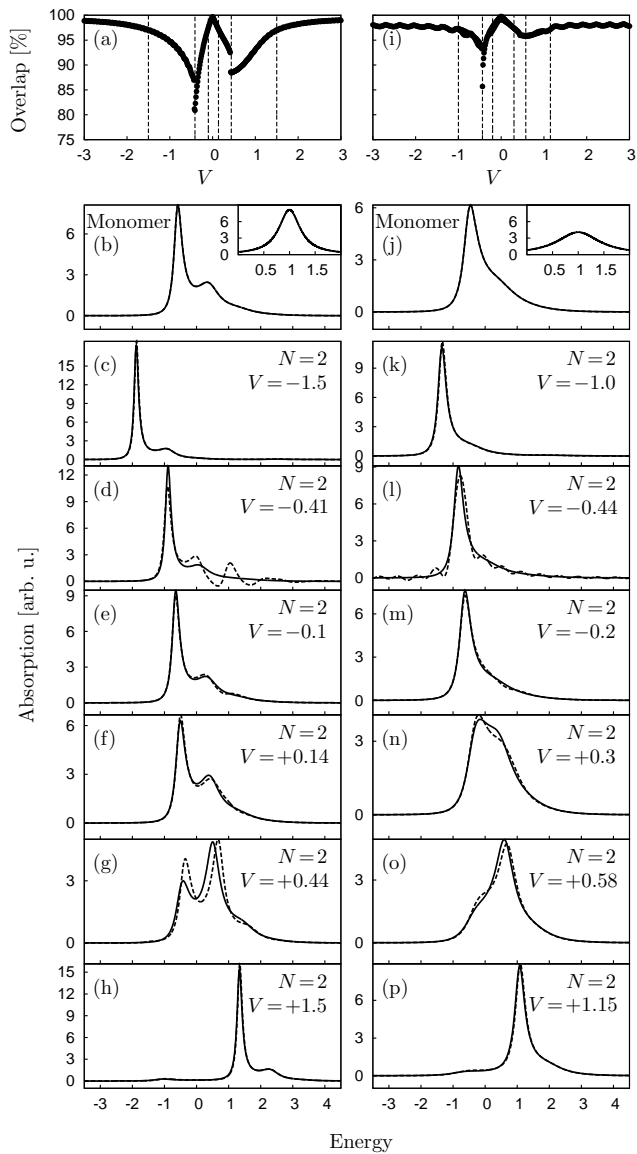}
  \caption{\label{fig1_spec_m1n2_om1x0.64_gam0.25_gam0.5} Left column: (a) Overlap of ZOFE spectrum with PM spectrum over $V$ for a dimer for a single Lorentzian spectral density centered at $\Omega$ with $X=0.64$ and width $\gamma=0.25\,\Omega$. (b) Monomer spectrum and spectral density $J(\omega)$ in units of $0.1\,\hbar^2\Omega$ (inset). (c)-(h) Corresponding dimer ZOFE spectra (dashed) and PM spectra (solid) for the values of $V$ indicated by the dashed vertical lines in (a).
Right column: same as left column, but with $\gamma=0.5\,\Omega$.}
\end{figure}

First, we consider a spectral density that is a single Lorentzian centered at a frequency $\Omega$.
Here and in the following, we choose $\hbar\Omega$ as the unit of energy.
Consequently, we express $\gamma_j$ and $\Omega_j$ in units of $\Omega$ and $V$ is given in units of $\hbar\Omega$.
As a criterion for the agreement between a ZOFE spectrum and the corresponding PM spectrum we use the overlap of the areas of the two spectra.
An overlap of $100\%$ then means, that the two spectra are in perfect agreement.
In Figure~\ref{fig1_spec_m1n2_om1x0.64_gam0.25_gam0.5}a this overlap is plotted against the monomer-monomer interaction $V$ for the case of a dimer ($N=2$) for a single Lorentzian spectral density with Huang-Rhys factor $X=0.64$ and width $\gamma=0.25$.
From this plot we see, that the overlap has minima at roughly $V=\pm 0.4$. 
It increases rapidly going from there to smaller $|V|$ and increases (more slowly) going to larger $|V|$.
At $V=0$, the case of independent monomers, the overlap reaches $100\%$, since in this case the ZOFE is no approximation, but gives the exact result (see Appendix~\ref{ap_non_interac_monomers}).
This can also be seen in Figure~\ref{fig1_spec_m1n2_om1x0.64_gam0.25_gam0.5}b, where the ZOFE (dashed line) and PM (solid line) monomer spectrum ($V=0$) are shown for the considered spectral density $J(\omega)$ displayed in the inset of the figure.
There, the dashed and the solid line are indistinguishable, since they lie exactly on top of each other, showing the perfect agreement between ZOFE and PM monomer spectrum.
The monomer spectrum consists of broadened peaks separated by the vibrational energy $\hbar\Omega$ of the PM.
We have chosen the zero of the energy axis to be located at the mean of the monomer spectrum.
The values of the overlap in Figure~\ref{fig1_spec_m1n2_om1x0.64_gam0.25_gam0.5}a converge to $100\%$ also for large $|V|$, showing that in the case of strong monomer-monomer interaction the ZOFE approximation becomes accurate too.
In Figure~\ref{fig1_spec_m1n2_om1x0.64_gam0.25_gam0.5}a the values of $V$, for which the overlap is minimal, are indicated by two dashed vertical lines.
Further vertical lines mark the four values of $V$, where the overlap reaches $97\%$.  
The corresponding dimer ZOFE (dashed line) and PM (solid line) spectra for the marked values of $V$ are shown in Figure~\ref{fig1_spec_m1n2_om1x0.64_gam0.25_gam0.5}c-h.
Figure~\ref{fig1_spec_m1n2_om1x0.64_gam0.25_gam0.5}c shows the ZOFE and PM spectrum for $V=-1.5$, having an overlap of $97\%$ corresponding to the left most vertical line in Figure~\ref{fig1_spec_m1n2_om1x0.64_gam0.25_gam0.5}a.
As we can see here, this overlap value of $97\%$ represents nearly perfect agreement between the spectra. 
The spectrum is much narrower than the monomer spectrum (note the different scales of the absorption axes).
This narrowing in the case of strong interaction $V$ (usually termed exchange
narrowing) has also appeared in the investigation of Gaussian
disorder~\cite{Kn84_73_,WaEiBr08_044505_,FiKnWi91_7880_}, single vibrational modes~\cite{WaEiBr08_044505_} and in
semi-empirical theories \cite{EiBr06_376_,EiBr07_354_}.
Apparently, as for the monomer, in this case of strong negative $V$ the ZOFE approximation is quite accurate. 
Upon increasing $V$, one enters the intermediate interaction regime
where discrepancies between ZOFE and PM spectra appear.
At $V=-0.41$, where the overlap has a pronounced dip (see vertical line in
Figure~\ref{fig1_spec_m1n2_om1x0.64_gam0.25_gam0.5}a), the agreement between
the spectra is worst, shown in
Figure~\ref{fig1_spec_m1n2_om1x0.64_gam0.25_gam0.5}d.
However, when $V$ is slightly changed from this value the agreement increases rapidly.
Upon increasing $V$, at $V=-0.1$ the overlap reaches again $97\%$ (see Figure~\ref{fig1_spec_m1n2_om1x0.64_gam0.25_gam0.5}e) owing to the fact, that in the region of weak inter-monomer interaction, where the spectra are similar to the monomer spectrum, the ZOFE approximation gives again accurate results.
This also holds true for the case of positive weak interaction, as is demonstrated in Figure~\ref{fig1_spec_m1n2_om1x0.64_gam0.25_gam0.5}f and as can be seen from the overlap values in Figure~\ref{fig1_spec_m1n2_om1x0.64_gam0.25_gam0.5}a.
Increasing $V$ further to positive intermediate $V$ again leads to discrepancies between the spectra as in the case of negative intermediate $V$.
However, for positive $V$ the largest deviation between ZOFE and PM spectrum at $V=+0.44$ (see Figure~\ref{fig1_spec_m1n2_om1x0.64_gam0.25_gam0.5}g), where the overlap is $88\%$, is not as large as the deviation at the overlap minimum for negative $V$. 
For strong positive interaction, there is again perfect agreement between ZOFE and PM spectra, as is shown in Fig.~\ref{fig1_spec_m1n2_om1x0.64_gam0.25_gam0.5}h for $V=+1.5$.
These spectra have a strong blue shift w.r.t.\ the monomer spectrum and have become narrower again.

For a larger $\gamma$, which leads to a faster decay of the bath correlation
function $\alpha(\tau)$ (see Eq.~(\ref{BathCorrFuncForLorentzSpecDens})), we expect that the agreement between ZOFE and PM spectra becomes better, since for infinitely fast decay (the Markov limit) ZOFE is exact (see Appendix~\ref{ap_markovian_bath}).
That this reasoning is indeed correct is demonstrated in the right column of
Fig.~\ref{fig1_spec_m1n2_om1x0.64_gam0.25_gam0.5}, where $\gamma=0.5$,
i.e.\ twice as large as in the left column. 
In the right column, the minimum values of the overlap of ZOFE and PM spectra are not as small as in the left column and the deviations between the spectra overall have become smaller compared to the left column.

\begin{figure}[h]
  \includegraphics[width=1\mylenunit]{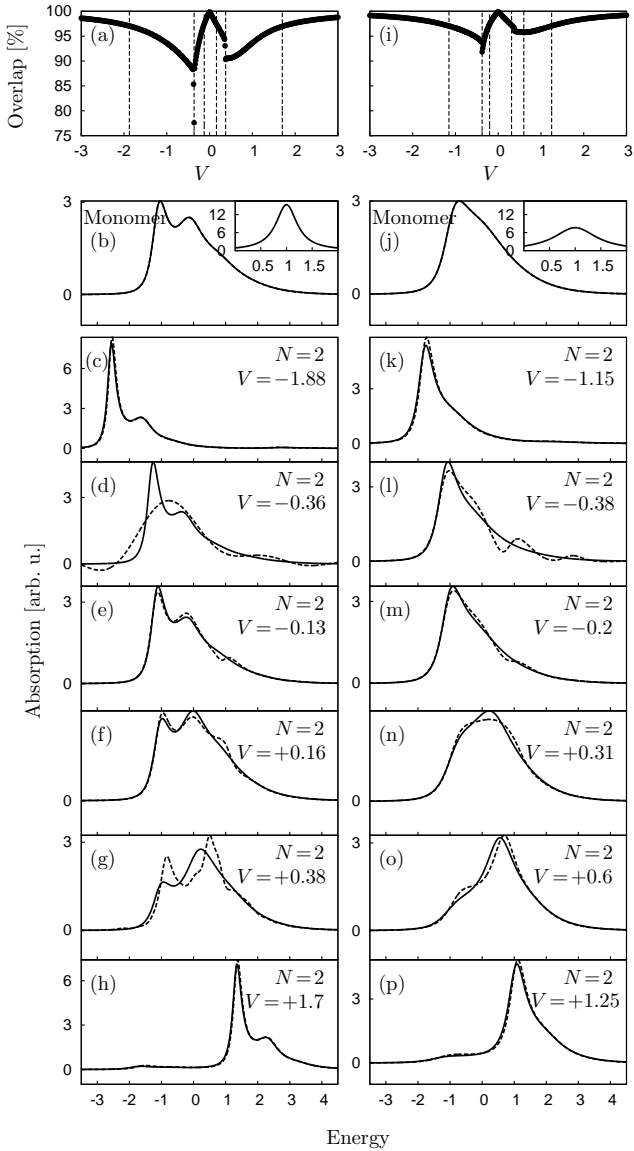}
  \caption{\label{fig2_spec_m1n2_om1x1.2_gam0.25_gam0.5} Same as
    Fig.~\ref{fig1_spec_m1n2_om1x0.64_gam0.25_gam0.5}, but with $X=1.2$.}
\end{figure}

Next, we consider the dependence of the agreement between the spectra on the Huang-Rhys factor $X$ (and thus on the coupling strength $\Gamma$).
In Fig.~\ref{fig2_spec_m1n2_om1x1.2_gam0.25_gam0.5}, overlap values and spectra are shown for a $X$ roughly twice as large as in  Fig.~\ref{fig1_spec_m1n2_om1x0.64_gam0.25_gam0.5}; the values of all other parameters are the same.
The stronger coupling of the PM to the electronic excitation can clearly be
seen in the monomer spectra
(Fig.~\ref{fig2_spec_m1n2_om1x1.2_gam0.25_gam0.5}b,~j) by an increase of the
intensity of the second peak (located roughly at energy $0$).
The overlap of ZOFE and PM spectra in Fig.~\ref{fig2_spec_m1n2_om1x1.2_gam0.25_gam0.5}a shows a similar behaviour as before, but now the minimum has a very singular character for $\gamma=0.25$.
The agreement between ZOFE and PM spectra at the overlap minimum (see Fig.~\ref{fig2_spec_m1n2_om1x1.2_gam0.25_gam0.5}d) becomes worse compared to the case of $X=0.64$ in Fig.~\ref{fig1_spec_m1n2_om1x0.64_gam0.25_gam0.5}d.
However, for $\gamma=0.5$ the largest deviation of the $X=1.2$ spectra is smaller than the largest deviation of the $X=0.64$ spectra.
Comparing the overlap plots for $X=0.64$ and $X=1.2$ (Fig.~\ref{fig1_spec_m1n2_om1x0.64_gam0.25_gam0.5}a,~i and Fig.~\ref{fig2_spec_m1n2_om1x1.2_gam0.25_gam0.5}a,~i), one sees that the agreement between ZOFE and PM spectra in the case of $X=0.64$ increases faster when going from the overlap minima to larger $|V|$ than in the case of $X=1.2$ (for both values of $\gamma$).
This is also reflected in the larger values of $|V|$ to which one has to go in the case of $X=1.2$ compared to $X=0.64$ (for the respective $\gamma$), to achieve perfect agreement between the spectra (see Fig.~\ref{fig1_spec_m1n2_om1x0.64_gam0.25_gam0.5}c,~h,~k,~p and Fig.~\ref{fig2_spec_m1n2_om1x1.2_gam0.25_gam0.5}c,~h,~k,~p).
These observations show that the quality of the ZOFE approximation depends on the magnitude of the inter-monomer interaction $V$ relative to the magnitude of the coupling $\Gamma=(\hbar\Omega)^2 X$ of the electronic excitation to the PM.

\begin{figure}[h]
  \includegraphics[width=1\mylenunit]{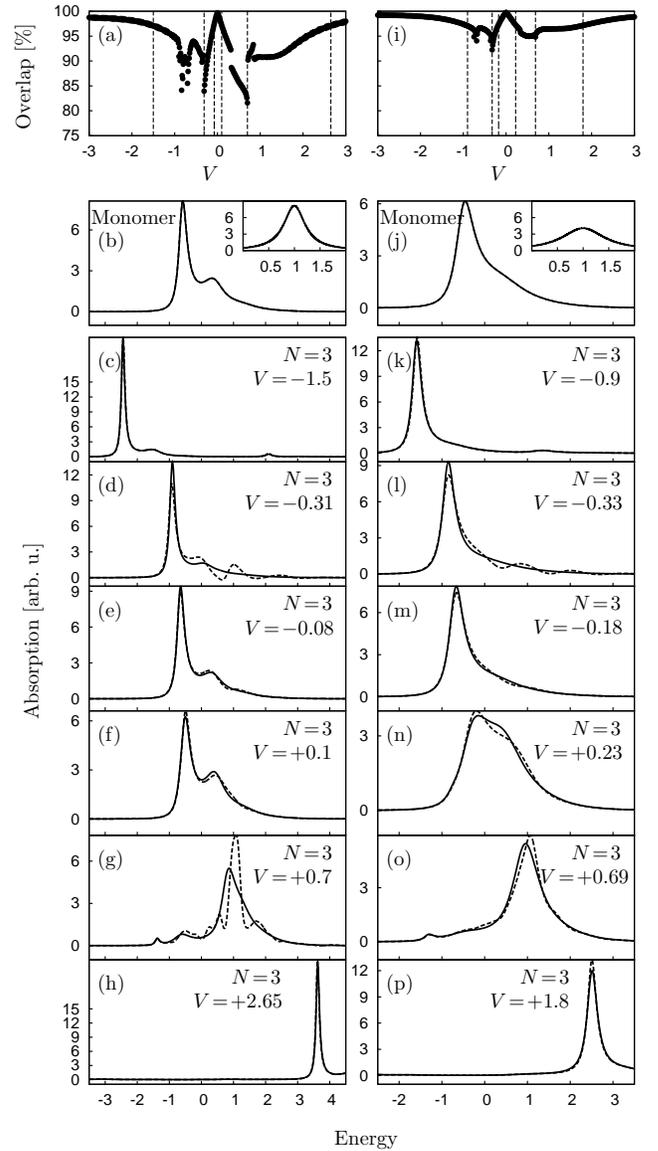}
  \caption{\label{fig3_spec_m1n3_om1x0.64_gam0.25_gam0.5} Same as
    Fig.~\ref{fig1_spec_m1n2_om1x0.64_gam0.25_gam0.5}, but for the case of a
    trimer ($N=3$).}
\end{figure}

In Figure~\ref{fig3_spec_m1n3_om1x0.64_gam0.25_gam0.5}, overlap values and absorption spectra for the same spectral densities as in Figure~\ref{fig1_spec_m1n2_om1x0.64_gam0.25_gam0.5} are shown, but now for a trimer ($N=3$).
Here, the plots of the overlap against $V$ show a greater number of local extrema.
The trimer spectrum for the strong negative coupling $V=-1.5$ looks similar to the respective dimer spectrum, but is slightly narrower and has an additional small peak located roughly at an energy of $+2.1$, due to a second allowed electronic transition.
In this case, ZOFE and PM spectrum agree perfectly.
Upon increasing $V$, as for the dimer we can identify the cases of strong and weak negative and positive interaction $V$, where there is good agreement between the spectra and the cases of negative and positive intermediate $V$, where the ZOFE and PM spectra show deviations.
However, for the trimer we observe, that in the region of strong positive $V$ to achieve perfect agreement between the spectra, we have to go to larger $V$, namely $V\approx +2.7$, compared to $V\approx +1.5$ for the dimer.
Broadening of the spectral density by a factor of two (see right column) leads for the trimer spectra basically to the same effects and to the better agreement between solid and dashed spectra as for the dimer. 

Our findings from the comparison of ZOFE spectra with exact PM spectra for the case of a spectral density with a single Lorentzian can be summarized as follows:
1) We always observe perfect agreement between ZOFE and PM spectra for strong and weak negative and positive inter-monomer interaction $V$.\\
2) There are deviations between the spectra in the intermediate $V$ region
(and clear resonance-like local minima in the overlap plots), that become smaller upon increasing the width $\gamma$ of the spectral density.\\
3) Increasing the coupling $\Gamma$ of the electronic excitation to the PM leads to a slower ascent of the agreement between the spectra when going from intermediate $|V|$ to larger $|V|$.\\
4) For the trimer, we find more local minima in the plots of the overlap against $V$ than for the dimer.\\
5) To achieve perfect agreement between the spectra in the region of strong positive interaction $V$, in the case of the trimer $V$ has to be larger than in the case of the dimer.

\begin{figure}[h]
  \includegraphics[width=1\mylenunit]{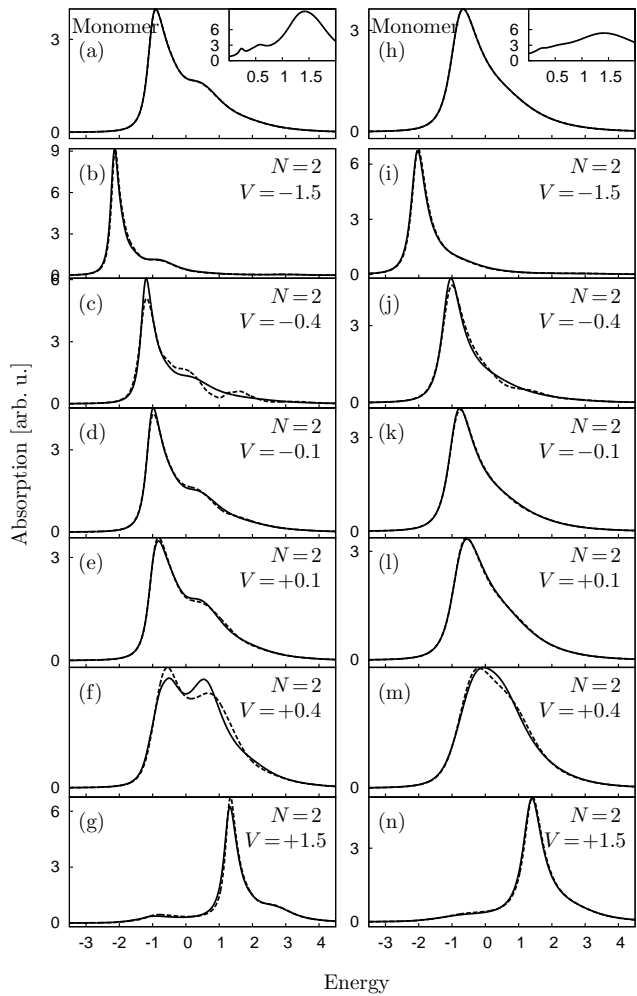}
  \caption{\label{fig4_spec_m6a_n2_gam0.25_gam0.5} Left column: (a) monomer spectrum for spectral density (shown in inset) that is a sum of six Lorentzians centered at different frequencies $\Omega_j=0.23, 0.42, 0.57, 1.29, 1.41, 1.61$ with different Huang-Rhys factors $X_j=0.4, 0.07, 0.18, 0.24, 0.12, 0.24$, where the width of each Lorentzian is $\gamma_j=0.25\,\Omega_j$.
(b)-(g) Corresponding dimer ZOFE spectra (dashed) and PM spectra (solid) for different monomer-monomer interaction $V$.
Right column: same as left column, but with $\gamma_j=0.5\,\Omega_j$ for each Lorentzian of the spectral density.}
\end{figure}

We have also performed exact calculations for more complex spectral densities, that are the sum of several Lorentzians.
Here, we have observed similar trends as described above for the case of the single Lorentzian spectral density.
As an example, in  Fig.~\ref{fig4_spec_m6a_n2_gam0.25_gam0.5} ZOFE and PM monomer and dimer spectra are shown for a spectral density, that is the sum of six Lorentzians centered at different frequencies $\Omega_j$ with different Huang-Rhys factors $X_j$ and widths $\gamma_j$. 
The six frequencies, where the Lorentzians are centered, range from $\Omega_1=0.23$ to $\Omega_6=1.61$ and the corresponding Huang-Rhys factors have different values $X_j=0.4\dots 0.24$.
In the left column of Fig.~\ref{fig4_spec_m6a_n2_gam0.25_gam0.5}, each Lorentzian has a width $\gamma_j=0.25\,\Omega_j$ and in the right column this width is increased by a factor two, i.e.\  $\gamma_j=0.5\,\Omega_j$.
The calculation of a ZOFE dimer or trimer spectrum for the single Lorentzian spectral densities of the Figures~\ref{fig1_spec_m1n2_om1x0.64_gam0.25_gam0.5}-\ref{fig3_spec_m1n3_om1x0.64_gam0.25_gam0.5}, as well as for the spectral densities of Fig.~\ref{fig4_spec_m6a_n2_gam0.25_gam0.5} consisting of six Lorentzians, can be done within a few seconds on a standard PC.
However, while the calculation of a PM dimer or trimer spectrum for the single Lorentzian spectral densities considered above also takes only a few seconds, the calculation of a PM dimer spectrum for the spectral densities consisting of six Lorentzians considered here took about six hours.
This increase of the computational effort makes investigations involving hundreds of spectra like the evaluation of the overlap against $V$ in Figures~\ref{fig1_spec_m1n2_om1x0.64_gam0.25_gam0.5}-\ref{fig3_spec_m1n3_om1x0.64_gam0.25_gam0.5} very time-consuming.
Therefore, in Fig.~\ref{fig4_spec_m6a_n2_gam0.25_gam0.5}  we simply choose values for the interaction $V$, that in the case of a single Lorentzian spectral density considered before are typical for the $V$ regions of perfect agreement and deviations between ZOFE and PM spectra. 
We see from Fig.~\ref{fig4_spec_m6a_n2_gam0.25_gam0.5}, that as for the single Lorentzian spectral densities also in this case of a more complex spectral density we obtain perfect agreement between ZOFE and PM spectra for strong and weak negative and positive interaction $V$ and deviations in the intermediate negative and positive $V$ regions.   
And also here, the deviations become smaller when we go from the left column of Fig.~\ref{fig4_spec_m6a_n2_gam0.25_gam0.5} to the right column, i.e.\ increase the width of the Lorentzians of the spectral density from $\gamma_j=0.25\,\Omega_j$ to $\gamma_j=0.5\,\Omega_j$.


\section{Summary and conclusions}
\label{conclusion}

In this work we presented two different methods to calculate zero temperature absorption
spectra of molecular aggregates, suited for situations in which the electronic excitation of a monomer
couples to a structured phonon environment.
Both methods are based on an open system approach, where the full Hamiltonian
is divided into a ``system'' part, an ``environment'' and the interaction
between them.

In the first method, we use the non-Markovian quantum state diffusion (NMQSD)
approach, where the system contains only electronic degrees of
  freedom. To tackle functional derivatives that appear in the NMQSD evolution
  equation the zeroth order functional expansion (ZOFE) approximation has been
utilized.
Although the NMQSD is based on a stochastic Schr\"odinger equation, we have
shown that in the calculation of the zero temperature absorption spectrum
the stochastic processes do not enter explicitly, so that only a
  deterministic equation has to be solved.
Since in this method  the system part contains only electronic degrees of freedom, i.e.\ the dimension of the considered single-exciton Hamiltonian is $N$ for an $N$-mer, this approach can be applied to quite large systems with complex structured environments.
The absorption spectrum of a 15-mer with a highly complex spectral density of the environmental modes for instance can be calculated within several minutes on a standard PC. 

To calculate the absorption spectrum exactly and find out the accuracy of the ZOFE approximation used by the first method, we applied a second method, the method of pseudomodes (PM) described in section~\ref{sec:Exact_solv_model}. 
Here, vibrational modes (pseudomodes) are included into the system part, which couple to the electronic excitation.
The electronic excitation now is not coupled directly to an environment anymore, but each PM couples to its own Markovian bath of vibrational modes.
This allows for a description of the problem via a Markovian quantum state diffusion (QSD) evolution equation, that can be solved numerically exact.
For the practically important case of a spectral density of the environmental modes that is a sum of Lorentzians~\cite{MeTa99_3365_,ScKlSc06_084903_,Ga97_2290_,Im94_3650_} the absorption spectrum of the first method without applying the ZOFE approximation is equal to the absorption spectrum provided by the PM method (as we show in Appendix~\ref{sec:ap_equiv_models_with_and_without_pm}). 
However, in the PM method the price one has to pay is that for each Lorentzian in the spectral density one PM is included into the system part, leading to a rapid growth of the dimension of the system Hamiltonian as the number $N$ of monomers or the number of Lorentzians in the spectral density is increased.
Thus, using the PM method we are at most able to calculate absorption spectra for a dimer with a spectral density of about six Lorentzians or a trimer with about five Lorentzians etc.\ due to limited computer capabilities. 
Here, the use of the Markovian QSD is advantageous over a propagation of a density matrix, having a size that is the square of the dimension of the used basis.
In the QSD, for the calculation of the zero temperature absorption spectrum, only a system of differential equations with the size of the basis has to be solved. 
The spectrum is then found by a simple Fourier transformation. 
For our choice of the basis, the Hamiltonian becomes very sparse, allowing an efficient numerical propagation.

To investigate the applicability of the first method, that uses the ZOFE approximation, in section~\ref{sec:compar_nmqsd_pm} we compared the calculated spectra
with spectra calculated using the PM method for small aggregates consisting of $N=2$ or $N=3$ monomers
for spectral densities consisting of one and six Lorentzians.
We always found perfect agreement between ZOFE and PM spectra in the cases of strong or weak inter-monomer interaction (that is, when the interaction energy is large or small compared to the width of the monomer spectrum).
However, we observed deviations between the spectra in the intermediate interaction regime, but these deviations become small upon increasing the width of the spectral density.
In particular, for some narrow regions of the interaction energy, where we found a resonance-like decrease of the agreement between the spectra, the deviations became relevant.
But also for these values of the interaction energy, the agreement becomes good again when the width of the spectral density is increased.
When going from intermediate to strong inter-monomer interaction $V$, the agreement between the spectra increases.
We found that in order to obtain the same degree of agreement for a stronger coupling $\Gamma$ of the electronic excitation to the vibrations, $V$ has to be chosen larger.
For more complex spectral densities consisting of several Lorentzians we basically observed the same trends as for a single Lorentzian. 
Let us finally stress that over the entire range of the inter-monomer interaction $V$, the NMQSD-ZOFE approach leads to spectra with an overlap of more than 80--90\% with the exact spectra, as can be seen in Figs.~\ref{fig1_spec_m1n2_om1x0.64_gam0.25_gam0.5}--\ref{fig3_spec_m1n3_om1x0.64_gam0.25_gam0.5}.
 
The NMQSD approach with the ZOFE approximation seems to be well suited to
bridge the gap between the dimer case, which can be treated numerically exact
for a large range of spectral densities (e.g.\ via the PM approach) and very
large aggregates where semi-empirical approximations (like the coherent
exciton scattering (CES) approximation\cite{EiBr06_376_,EiBr07_354_,EiKnBr07_104904_}) lead to excellent agreement between theory and experiment.  

As  shown in Ref.~\citep{RoEiWo09_058301_}, the  NMQSD-ZOFE approach  is not restricted to the calculation of absorption spectra, but allows also the calculation of fully time dependent quantities like the probability to find excitation on a certain monomer. 
  Then however, in contrast to the present situation, stochastic processes will explicitly enter into the calculations.

\appendix

\section{Exactly solvable cases}
\label{ap_exact_solvable}
\subsection{Non-interacting monomers}
\label{ap_non_interac_monomers}

Here we demonstrate that ZOFE leads to the exact spectrum for non-interacting
monomers.
To this end we show that the approximate zeroth order operator $O_0^{(n)}(t,s)$, which
appears within ZOFE (see Eq.~(\ref{NoNoiseApprox})), satisfies the fundamental consistency condition
Eq.~(\ref{consis_condition}) or equivalently the evolution equation
Eq.~(\ref{EvolEqOOpWithFuncDer}):
 
We insert  $O_0^{(n)}(t,s)$ into Eq.~(\ref{EvolEqOOpWithFuncDer}).
For non-interacting monomers, i.e.\ for $V_{nm}=0$, we find
\begin{equation}
\label{EvolEqOOpMon}
  \partial_t O_0^{(n)}(t,s)=0,
\end{equation}  
since the commutators vanish and the functional derivative yields zero.
Equation~(\ref{EvolEqOOpMon}) together with the initial condition
$O_0^{(n)}(s,s)=L_n$ (Eq.~(\ref{init_cond_OOp})) yields
the constant
\begin{equation}
\label{const_OOp_Mon}
  O_0^{(n)}(t,s)=L_n
\end{equation}
without any approximation. Thus, ZOFE is exact for non-interacting monomers.

\subsection{Markovian environment}
\label{ap_markovian_bath}

Within the QSD approach of section~\ref{sec:qsd_approach} we choose the bath correlation 
function $\alpha_n(t-s)$ to be delta-like, i.e.\
\begin{equation}
\label{MarkovianBathCorrFctQSD}
  \alpha_n(t-s)\equiv 2\hbar^2\theta_n\delta(t-s)
\end{equation}
which amounts to a Markovian environment.
Inserting Eqs.~(\ref{MarkovianBathCorrFctQSD}) and~(\ref{init_cond_OOp}) 
into Eq.~(\ref{DefOQuer}) yields
\begin{equation}
\label{OQuerForMarkovianBath}
  \bar{O}^{(n)}(t,{\mathbf z}^*)=\theta_nL_n.
\end{equation}
We see that in this Markov limit, the exact operator $\bar{O}^{(n)}(t,{\mathbf z}^*)$ 
is independent of ${\mathbf z}^*$ and constant w.r.t.\ $t$. Thus, we find that ZOFE also
contains the Markov limit.
We insert Eq.~(\ref{OQuerForMarkovianBath}) into Eq.~(\ref{EvolEqPsiSysSpaceMitOQuer}) 
and obtain the evolution equation
\begin{equation}
\label{EvolEqPsiSysSpaceMitOQuerMarkovBath}
\begin{split}
  \partial_t&\ket{\psi(t,{\mathbf z}^*)}=-\frac{i}{\hbar} H_{\rm sys}\ket{\psi(t,{\mathbf z}^*)}\\
&+\sum_n\left(L_nz^*_{t,n}-\theta_n L^{\dagger}_n L_n  \right)\ket{\psi(t,{\mathbf z}^*)}.
\end{split}
\end{equation}
which is the well-known linear Markov QSD equation~\cite{Pe98__}.

\section{Absorption in the pseudomode approach} 
\label{sec:ap_absorpPM}

In this section, we show how we obtain the absorption spectrum using the approach of pseudomodes of Section~\ref{sec:Exact_solv_model}.
As the initial state of the aggregate we take
\begin{equation}
  \label{InitAggStateAbsPM}
  \ket{\tPhi(t=0)}=\ket{g_{\rm el}}\ket{g_{\rm PM}}\ket{\tNull},
\end{equation}
where $\ket{g_{\rm el}}$ is the product of all monomer electronic ground states given by Eq.~(\ref{ElGroundStateAgg}).
The state $\ket{g_{\rm PM}}$ denotes the product of the ground states of all PMs 
and $\ket{\tNull}$ is the bath state in which all bath modes are in their ground states.
Analogously to Section~\ref{sec:AbsOfAgg}, the cross section for absorption of light can be obtained from Eq.~(\ref{AbsSpecPM}) with the dipole correlation function
\begin{equation}
\label{GeneralTimeCorrFctPM}
  \tM(t)=\bra{\tPhi(t=0)}\hat{\vec{\mu}}\cdot\vec{\mathcal{E}}\ e^{-i\tH t/\hbar}\ \hat{\vec{\mu}}\cdot\vec{\mathcal{E}^*}\ket{\tPhi(t=0)},
\end{equation}
where $\hat{\vec{\mu}}$ is the total dipole operator of the aggregate, given by Eq.~(\ref{TotalDipSumOfMonDip}).
As in Section~\ref{sec:AbsOfAgg} by using 
Eqs.~(\ref{TotalDipSumOfMonDip}) and (\ref{MonTransDip}), we obtain
\begin{equation}
\label{MofTwithMuPrefactorPM}
  \tM(t)=\mu_{\rm tot}^2\braket{\tPsi(t=0)}{\tPsi(t)}
\end{equation}
with the normalized aggregate states
\begin{equation}
\label{DefPsiTimeZeroPM}
  \ket{\tPsi(t=0)}=\ket{\tpsi_0}\ket{\tNull}
\end{equation}
and with $\ket{\tpsi_0}$ given by Eq.~(\ref{DefPsiNullPM}) and 
\begin{equation}
\label{PsiOftPropagatorPM}
  \ket{\tPsi(t)}=\exp(-i\tH^e t/\hbar)\ket{\tPsi(t=0)}.
\end{equation}
The prefactor $\mu_{\rm tot}$ in Eq.~(\ref{MofTwithMuPrefactorPM}) is defined in Eq.~(\ref{DefMuPrefac}).
Similarly as in Section~\ref{sec:abs_in_qsd}, we use the expansion 
\begin{equation}
\label{ExpandTotalPsiWithBargmannPM}
  \ket{\tPsi(t)}=\int\frac{d^2{\mathbf \tz}}{\pi}e^{-|{\mathbf \tz}|^2}\ket{\tpsi(t,{\mathbf \tz}^*)}\ket{{\mathbf \tz}}.
\end{equation}
Here, $\ket{\tz_{nj\rho}}=\exp(\tz_{nj\rho}\ta^{\dagger}_{nj\rho})\ket{\tNull_{nj\rho}}$
denote Bargmann coherent bath states, 
and $\ket{{\mathbf \tz}}=\prod_n\prod_j\prod_{\rho}\ket{\tz_{nj\rho}}$.
From Eqs.~(\ref{DefPsiTimeZeroPM}), 
(\ref{ExpandTotalPsiWithBargmannPM}), and (\ref{MofTwithMuPrefactorPM}), we obtain
\begin{equation}
\label{M_of_t_QSD_PM_z_zero}
  \tM(t)=\mu_{\rm tot}^2\braket{\tpsi_0}{\tpsi(t,{\mathbf \tz}^*=0)}.
\end{equation}

We may determine the state $\ket{\tpsi(t,{\mathbf \tz}^*=0)}$
using the QSD approach analogously to Section~\ref{sec:qsd_approach},
now for a memory-less (Markovian) environment (see also appendix \ref{ap_markovian_bath}).
First, $\tH^e$ of Eq.~(\ref{H_e_PM}) is transformed to the interaction 
representation w.r.t.\ $\tH_{\rm env}$ Eq.~(\ref{H_EnvPM}) to find
\begin{equation}
\label{H_e_Interac_represPM}
  \tH^e(t)=\tH_{\rm sys}+\sum_{n=1}^N\sum_j\left(\tL_{nj}\tA^{\dagger}_{nj}(t)+\tL^{\dagger}_{nj}\tA_{nj}(t)\right)
\end{equation}
with 
\begin{equation}
\label{Def_aSchlangeVonT}
  \tA_{nj}(t)\equiv \sum_{\rho}\tkappa_{nj\rho}^* e^{-i\tomega_{nj\rho}t}\ta_{nj\rho}
\end{equation}
and where
\begin{equation}
  \tL_{nj}\equiv b_{nj}.
\end{equation}
From the definition of $\tA_{nj}(t)$ Eq.~(\ref{Def_aSchlangeVonT}), we get the bath correlation function
\begin{equation}
\label{BathCorrFctAlphaSchlangePM}
\begin{split}
  \talpha_{nj}(t-s)&\equiv \bra{\tNull}\tA_{nj}(t)\tA_{nj}^{\dagger}(s)\ket{\tNull}\\
&=\sum_{\rho}|\tkappa_{nj\rho}|^2 e^{-i\tomega_{nj\rho}(t-s)},
\end{split}
\end{equation}
where $\ket{\tNull}$ denotes the state in which all bath modes are in their ground states.

We want to solve the Schr\"odinger equation 
\begin{equation}
\label{SchroeEqPsiTotalPM}
  i\hbar\partial_t\ket{\tPsi(t)}=\tH^e(t)\ket{\tPsi(t)}.
\end{equation}
Inserting the expansion Eq.~(\ref{ExpandTotalPsiWithBargmannPM}) into Eq.~(\ref{SchroeEqPsiTotalPM}) we obtain
\begin{equation}
\label{EvolEqPsiSysSpacePM}
\begin{split}
\partial_t \ket{\tpsi(t,{\mathbf \tz}^*)}
&=
-\frac{i}{\hbar} \tH_{\rm sys}\ket{\tpsi(t,{\mathbf \tz}^*)}
 +\sum_{n,j} \tL_{nj} \tz^*_{t,nj}\ket{\tpsi(t,{\mathbf \tz}^*)}\\
& -\frac{1}{\hbar^2}\sum_{n,j} \tL_{nj}^{\dagger} \int_0^t ds\ \talpha_{nj}(t-s) \frac{\delta \ket{\tpsi(t,{\mathbf \tz}^*)} }{\delta \tz^*_{s,nj} }
\end{split}
\end{equation} 
with time-dependent complex numbers
\begin{equation}
  \tz^*_{t,nj}=-\frac{i}{\hbar}\sum_{\rho}\tkappa_{nj\rho}\tz^*_{nj\rho}e^{i\tomega_{nj\rho}t}.
\end{equation}

From Eq.~(\ref{MarkovBathCorrFuncPM}) we have $\talpha_{nj}(t-s) \propto \delta(t-s)$, 
i.e.\ the environment of the pseudomodes is Markovian.
Now, we make use of this fact by inserting Eq.~(\ref{MarkovBathCorrFuncPM}) into 
Eq.~(\ref{EvolEqPsiSysSpacePM}) and obtain
\begin{equation}
\label{EvolEqPsiSysSpacePMMarkovian}
\begin{split}
\partial_t \ket{\tpsi(t,{\mathbf \tz}^*)}
=&
-\frac{i}{\hbar} \tH_{\rm sys}\ket{\tpsi(t,{\mathbf \tz}^*)}\\
& +\sum_{n,j} \tL_{nj} \tz^*_{t,nj}\ket{\tpsi(t,{\mathbf \tz}^*)}\\
& -\sum_{n,j} \gamma_{nj}\tL_{nj}^{\dagger}\tL_{nj}\ket{\tpsi(t,{\mathbf \tz}^*)}.
\end{split}
\end{equation} 
In the case ${\mathbf \tz}^*=0$ equation~(\ref{EvolEqPsiSysSpacePMMarkovian}) yields the evolution equation~(\ref{EvolEqToCalcPMAbsSpec}) of Section~\ref{sec:ExactAbsSpec}, i.e.\ 
\begin{equation}
\label{EvolEqPsiSysSpacePMMarkovianZEqualZero}
\begin{split}
\partial_t \ket{\tpsi(t,{\mathbf \tz}^*=0)}
=&
-\frac{i}{\hbar} \tH_{\rm sys}\ket{\tpsi(t,{\mathbf \tz}^*=0)}\\
& -\sum_{n,j} \gamma_{nj}\tL_{nj}^{\dagger}\tL_{nj}\ket{\tpsi(t,{\mathbf \tz}^*=0)}.
\end{split}
\end{equation} 
In order to obtain the absorption spectrum, this equation is solved for the 
initial state $\ket{\tpsi_0}$.

\subsubsection{Numerical implementation}

We express
the Schr\"odinger equation Eq.~(\ref{EvolEqToCalcPMAbsSpec}) in a basis of product states 
\begin{equation}
\label{DefThetaStates}
  \ket{\theta^{\beta}_n}\equiv \ket{\pi_n}\prod_{n=1}^N\prod_j\ket{\beta_{nj}}
\end{equation}
with vibrational eigenstates $\ket{\beta_{nj}}$ of PM $j$ in the electronic ground state 
of monomer $n$, i.e. the states $\ket{\beta_{nj}}$ satisfy
\begin{equation}
  \hbar\Omega_{nj}b_{nj}^{\dagger}b_{nj}\ket{\beta_{nj}}=\hbar\Omega_{nj}\beta_{nj}\ket{\beta_{nj}}.
\end{equation}
From the Schr\"odinger equation Eq.~(\ref{EvolEqToCalcPMAbsSpec}) we obtain a system of coupled differential equations for the components $\braket{\theta_n^{\beta}}{\tpsi(t)}$ w.r.t.\ the states Eq.~(\ref{DefThetaStates})
\begin{equation}
\label{CoupDiffSchroeEqPM}
\begin{split}
   \partial_t&\braket{\theta_n^{\beta}}{\tpsi(t)} = \\
& -\frac{i}{\hbar}\left(\varepsilon_n+\sum_{m=1}^N\sum_j(\hbar\Omega_{mj}-i\hbar\gamma_{mj})\beta_{mj}\right)\braket{\theta_n^{\beta}}{\tpsi(t)}\\
& +\frac{i}{\hbar}\sum_j\sqrt{\Gamma_{nj}}\sqrt{\beta_{nj}}\braket{\theta_n^{(\beta_{11}\dots\beta_{nj}-1\dots)}}{\tpsi(t)}  \\
& +\frac{i}{\hbar}\sum_j\sqrt{\Gamma_{nj}}\sqrt{\beta_{nj}+1}\braket{\theta_n^{(\beta_{11}\dots\beta_{nj}+1\dots)}}{\tpsi(t)} \\
 & -\frac{i}{\hbar}\sum_{m\neq n}^N V_{nm}\braket{\theta_m^{\beta}}{\tpsi(t)}.
\end{split}
\end{equation}  
Since the corresponding matrix is very sparse, we are able to calculate the absorption spectrum 
taking into account a few PMs per monomer.
For the calculation of the dipole correlation function Eq.~(\ref{M_of_t_QSD_PM}), we 
choose the initial state Eq.~(\ref{DefPsiNullPM}) to be real-valued.
On that condition and by using the fact that the Hamiltonian corresponding to the system of 
equations Eq.~(\ref{CoupDiffSchroeEqPM}) is symmetric (it is equal to its transpose), 
we can calculate the value of the dipole correlation function at time $2t$ through~\cite{En92_76_}
\begin{equation}
\label{M_of_t_QSD_PM_double_time}
  \tM(2t)=\mu_{\rm tot}^2\left(\bra{\tpsi(t)}\right)^*\ket{\tpsi(t)},
\end{equation}
where the star $*$ denotes the complex conjugate.
That is, the time we need to propagate the wavefunction numerically shortens by a factor of two.
Note, that for complex initial wavefunctions the efficient calculation scheme 
Eq.~(\ref{M_of_t_QSD_PM_double_time}) is no longer applicable.

\section{Equality of NMQSD spectrum and PM spectrum}
\label{sec:ap_equiv_models_with_and_without_pm}

In this section, we show that the absorption spectrum obtained from the NMQSD approach 
is equal to the spectrum obtained from the PM method
for a bath correlation function consisting of a sum of exponentials as in
Eq.~(\ref{BathCorrFuncForLorentzSpecDens}).
This equivalence holds true provided
the parameters ($\Omega_{nj}, \gamma_{nj}$, and $\Gamma_{nj}$) of the pseudomode
description are taken from the corresponding
bath correlation function of the NMQSD approach.

We start with the Hamiltonian of the PM approach defined by Eqs.~(\ref{H_e_PM})-(\ref{H_IntPM}). 
It can be written as 
\begin{equation}
\label{HePMwithHvib}
  \tH^e=H_{\rm sys}+\sum_{n=1}^N\sum_j\sqrt{\Gamma_{nj}}\left(L_nb^{\dagger}_{nj}+L_n^{\dagger}b_{nj}  \right)+H_{\rm vib}.
\end{equation} 
Here, $H_{\rm sys}$ is the Hamiltonian containing only electronic degrees of freedom and
\begin{equation}
\begin{split}
\label{DefHvib}
   &H_{\rm vib}\equiv
   \sum_{n=1}^N\sum_j\hbar\Omega_{nj}b^{\dagger}_{nj}b_{nj}\\
&+\sum_{n=1}^N\sum_j\sum_{\rho}\left(\tkappa_{nj\rho}^*\ta_{nj\rho}b^{\dagger}_{nj}+\tkappa_{nj\rho}\ta_{nj\rho}^{\dagger}b_{nj}\right)\\
&+\sum_{n=1}^N\sum_j\sum_{\rho}\hbar\tomega_{nj\rho}\ta_{nj\rho}^{\dagger}\ta_{nj\rho},
\end{split}
\end{equation}
contains the pseudomodes, their Markovian environments, and their respective couplings.
We diagonalize $H_{\rm vib}$, i.e.\ we write
\begin{equation}
\label{DiagHvib}
  H_{\rm vib}=\sum_{n=1}^N\sum_j\sum_{\sigma}\hbar\homega_{nj\sigma}c^{\dagger}_{nj\sigma}c_{nj\sigma},
\end{equation}
with new frequencies $\homega_{nj\sigma}$ and new annihilation operators $c_{nj\sigma}$ that are a 
linear combination
\begin{equation}
\label{C_as_superp_of_b_and_a}
  c_{nj\sig}=\chi_{nj\sig}^*b_{nj}+\sum_{\rho}\zeta_{nj\rho\sig}\ta_{nj\rho}
\end{equation}
of the original ones and with
\begin{equation}
\label{b_as_superp_of_c}
  b_{nj}=\sum_{\sig}\chi_{nj\sig}c_{nj\sig},
\end{equation}
where the $\chi_{nj\sig}$ and $\zeta_{nj\rho\sig}$ are complex coefficients
{(this is shown in Ref.~\cite{BoFoNe06_472_} for real coefficients and can be
  extended easily to the present case)}.
With Eqs.~(\ref{DiagHvib}) and~(\ref{b_as_superp_of_c}) $\tH^e$ Eq.~(\ref{HePMwithHvib}) 
has the form of $H^e$ Eqs.~(\ref{HeAsSysPlusIntPlusEnv})-(\ref{HInt}) and we can derive a 
Schr\"odinger equation in the reduced space of $H_{\rm sys}$ analogously to 
section~\ref{sec:qsd_approach}.
Transforming $\tH^e$ from Eq.~(\ref{HePMwithHvib}) to the interaction 
representation w.r.t.\ $H_{\rm vib}$ yields
\begin{equation}
\label{H_e_Interac_represen_PM}
  \tH^e(t)=H_{\rm sys}+\sum_{n=1}^N\left(L_nB_n^{\dagger}(t)+L_n^{\dagger}B_n(t)\right),
\end{equation}
with
\begin{equation}
\label{DefBfromBSchlange}
  B_n(t)\equiv \sum_j\sqrt{\Gamma_{nj}}b_{nj}(t)
\end{equation}
and
\begin{equation}
\label{DefBSchlangeVonTime}
  b_{nj}(t)\equiv e^{iH_{\rm vib}t/\hbar}\ b_{nj}\  e^{-iH_{\rm vib}t/\hbar}.
\end{equation}
Note that Eq.~(\ref{H_e_Interac_represen_PM}) has the same structure as Eq.~(\ref{H_e_Interac_repres}).
Using Eqs.~(\ref{b_as_superp_of_c}) and~(\ref{DiagHvib}), from Eq.~(\ref{DefBSchlangeVonTime}) we obtain
\begin{equation}
\label{bOfTwithC}
  b_{nj}(t)=\sum_{\sig}\chi_{nj\sig}c_{nj\sig}e^{-i\homega_{nj\sig}t}.
\end{equation}
From Eqs.~(\ref{bOfTwithC}) and~(\ref{DefBfromBSchlange}), we get
\begin{equation}
  B_n(t)=\sum_j\sqrt{\Gamma_{nj}}\sum_{\sig}\chi_{nj\sig}c_{nj\sig}e^{-i\homega_{nj\sig}t}.
\end{equation}
Next we insert the expansion 
\begin{equation}
\label{ExpandTotalPsiWithBargmannPMElectr}
  \ket{\tPsi(t)}=\int\frac{d^2{\mathbf \hz}}{\pi}e^{-|{\mathbf \hz}|^2}\ket{\varphi(t,{\mathbf \hz}^*)}\ket{{\mathbf \hz}}
\end{equation}
w.r.t.\ Bargmann coherent states $\ket{\hz_{nj\sig}}=\exp(\hz_{nj\sig}c^{\dagger}_{nj\sig})\ket{\hxi^0_{nj\sig}}$, where $\ket{{\mathbf \hz}}=\prod_n\prod_j\prod_{\sig}\ket{\hz_{nj\sig}}$, into the Schr\"odinger equation
\begin{equation}
\label{SchroeEqPsiTotalPMElectr}
  i\hbar\partial_t\ket{\tPsi(t)}=\tH^e(t)\ket{\tPsi(t)}
\end{equation}
for the total state $\ket{\tPsi(t)}$.
This leads to an evolution equation
\begin{equation}
\label{EvolEqPsiSysSpacePMElectr}
\begin{split}
\partial_t \ket{\varphi(t,{\mathbf \hz}^*)}
&=
-\frac{i}{\hbar} H_{\rm sys}\ket{\varphi(t,{\mathbf \hz}^*)}
 +\sum_n L_n \hz^*_{t,n}\ket{\varphi(t,{\mathbf \hz}^*)}\\
& -\frac{1}{\hbar^2}\sum_n L_n^{\dagger} \int_0^t ds\ \beta_n(t-s) \frac{\delta \ket{\varphi(t,{\mathbf \hz}^*)} }{\delta \hz^*_{s,n} }
\end{split}
\end{equation} 
for the state $\ket{\varphi(t,{\mathbf \hz}^*)}$ in the Hilbert space of the
original ``system'' with Hamiltonian $H_{\rm sys}$.
In Eq.~(\ref{EvolEqPsiSysSpacePMElectr}), the time-dependent complex numbers
\begin{equation}
  \hz^*_{t,n}=-\frac{i}{\hbar}\sum_j\sqrt{\Gamma_{nj}}\sum_{\sig}\chi_{nj\sig}^*\hz_{nj\sig}e^{i\homega_{nj\sig}t}
\end{equation}
and the correlation functions
\begin{equation}
\label{bathCorrFuncTZeroPMElectr}
\begin{split}
  \beta_n(t-s)&=\left\langle B_n(t)B_n^{\dagger}(s)\right\rangle_{T=0}=\bra{\hNull} B_n(t)B_n^{\dagger}(s)\ket{\hNull}\\
&= \sum_j\Gamma_{nj}\sum_{\sig}|\chi_{nj\sig}|^2e^{-i\homega_{nj\sig}(t-s)}
\end{split}
\end{equation}
are used.

The NMQSD absorption spectrum is equal to the PM spectrum if the corresponding time correlation functions $M(t)$ of the NMQSD approach and $\tM(t)$ of the PM approach are equal (see Eqs.~(\ref{AbsSpec}) and~(\ref{AbsSpecPM})). 
From Eqs.~(\ref{MofTwithMuPrefactorPM})-(\ref{PsiOftPropagatorPM}) and Eq.~(\ref{ExpandTotalPsiWithBargmannPMElectr}) we get
\begin{equation}
\label{MPMofTwithIntBargmann}
  \tM(t)=\mu_{\rm tot}^2\bra{\tNull}\bra{g_{\rm PM}}\bra{\psi_0}\int\frac{d^2{\mathbf \hz }}{\pi}e^{-|{\mathbf \hz}|^2}\ket{\varphi(t,{\mathbf \hz}^*)}\ket{{\mathbf \hz}},
\end{equation}
similar to Eq.~(\ref{MofTwithIntBargmann}).
Because of equation~(\ref{C_as_superp_of_b_and_a}), we have
\begin{equation}
\label{EqualityOfNullStates}
  \ket{g_{\rm PM}}\ket{\tNull}=\ket{\hNull}.
\end{equation}
Using Eq.~(\ref{EqualityOfNullStates}), equation~(\ref{MPMofTwithIntBargmann}) yields
\begin{equation}
\label{M_of_t_QSD_PM_Electr}
  \tM(t)=\mu_{\rm tot}^2\braket{\psi_0}{\varphi(t,{\mathbf \hz}^*=0)}
\end{equation}
with the initial state
\begin{equation}
  \ket{\varphi(t=0,{\mathbf \hz}^*=0)}=\ket{\psi_0}.
\end{equation}
The time correlation function $\tM(t)$ Eq.~(\ref{M_of_t_QSD_PM_Electr}) of the PM method is 
equal to $M(t)$ Eq.~(\ref{M_of_t_QSD}) of the NMQSD approach, if the state 
$\ket{\varphi(t,{\mathbf \hz}^*=0)}$ is equal to $\ket{\psi(t,{\mathbf z}^*=0)}$ for all times $t$.
Thus, we next show that Eq.~(\ref{EvolEqPsiSysSpacePMElectr}) for ${\mathbf \hz}^*=0$ is 
equivalent to Eq.~(\ref{EvolEqPsiSysSpace}) for ${\mathbf z}^*=0$.
According to Eq.~(\ref{DefOfO}), we replace the functional derivative in Eq.~(\ref{EvolEqPsiSysSpace}) by the operator $O^{(n)}(t,s,{\mathbf z}^*)$ and the functional derivative in Eq.~(\ref{EvolEqPsiSysSpacePMElectr}) by an operator $\hO^{(n)}(t,s,{\mathbf \hz}^*)$. Note, however, that we do not make the ZOFE approximation, so that our treatment remains exact.
Following Ref.~\cite{YuDiGi99_91_}, one can expand the operators $O^{(n)}(t,s,{\mathbf z}^*)$ and $\hO^{(n)}(t,s,{\mathbf \hz}^*)$ w.r.t.\ the functionals $z^*_{t,n}$ and $\hz^*_{t,n}$ and obtain a hierarchy of differential equations for the different orders $O^{(n)}_0,O^{(n)}_1,\dots$ of the expansion.
These differential equations do not contain $z^*_{t,n}$ (and $\hz^*_{t,n}$ respectively) anymore and they are the same for  $O^{(n)}(t,s,{\mathbf z}^*)$ and $\hO^{(n)}(t,s,{\mathbf \hz}^*)$.
Thus we have $O^{(n)}_0=\hO^{(n)}_0$, $O^{(n)}_1=\hO^{(n)}_1,\dots$.
Now it only remains to show that in Eqs.~(\ref{EvolEqPsiSysSpace}) and~(\ref{EvolEqPsiSysSpacePMElectr}) $\alpha_n(t-s)=\beta_n(t-s)$.
This we will show in the following.
We consider the time derivative of Eq.~(\ref{DefBSchlangeVonTime})
\begin{equation}
\label{TimeDerivBSchlangeVonT}
  \partial_t b_{nj}(t)=\frac{i}{\hbar}e^{iH_{\rm vib}t/\hbar}\left[H_{\rm vib},b_{nj}\right]e^{-iH_{\rm vib}t/\hbar}.
\end{equation}   
Inserting the definition of $H_{\rm vib}$ Eq.~(\ref{DefHvib}) into Eq.~(\ref{TimeDerivBSchlangeVonT}) yields
\begin{equation}
\label{EvolEqBSchalngeVonT}
  \partial_t b_{nj}(t)=-i\Omega_{nj} b_{nj}(t)-\frac{i}{\hbar}\sum_{\rho}\tkappa_{nj\rho}^*\ta_{nj\rho}(t),
\end{equation}
with
\begin{equation}
\label{DefASchlangeVonT}
  \ta_{nj\rho}(t)\equiv e^{iH_{\rm vib}t/\hbar}\ \ta_{nj\rho}\  e^{-iH_{\rm vib}t/\hbar}.
\end{equation}
Taking the time derivative of Equation~(\ref{DefASchlangeVonT}) and inserting the definition of $H_{\rm vib}$ Eq.~(\ref{DefHvib}) yields
\begin{equation}
\label{EvolEqASchalngeVonT}
  \partial_t\ta_{nj\rho}(t)=-i\tomega_{nj\rho}\ta_{nj\rho}(t)-\frac{i}{\hbar}\tkappa_{nj\rho} b_{nj}(t).
\end{equation}
We integrate Eq.~(\ref{EvolEqASchalngeVonT}) and obtain
\begin{equation}
\label{IntegratedEvolEqASchalngeVonT}
\begin{split}
  \ta_{nj\rho}(t)=&e^{-i\tomega_{nj\rho}t}\ta_{nj\rho}(0)\\
&-\frac{i}{\hbar}\tkappa_{nj\rho}\int_0^t ds\ e^{-i\tomega_{nj\rho}(t-s)} b_{nj}(s).
\end{split}
\end{equation}
Inserting $\ta_{nj\rho}(t)$ Eq.~(\ref{IntegratedEvolEqASchalngeVonT}) into Eq.~(\ref{EvolEqBSchalngeVonT}) yields
\begin{equation}
\label{EvolEqForBSchlangeVonTWithBathCorrFct}
\begin{split}
  \partial_t b_{nj}(t)&=-i\Omega_{nj} b_{nj}(t)\\
&-\frac{1}{\hbar^2}\int_0^t ds\ \sum_{\rho}|\tkappa_{nj\rho}|^2 e^{-i\tomega_{nj\rho}(t-s)} b_{nj}(s)\\
&-\frac{i}{\hbar}\tA_{nj}(t)
\end{split}
\end{equation}
with $\tA_{nj}(t)$ given by Eq.~(\ref{Def_aSchlangeVonT}).
The $\tA_{nj}(t)$ are correlated with the function $\talpha_{nj}(t-s)$ given 
by Eq.~(\ref{BathCorrFctAlphaSchlangePM}).

Using the PM method the bath is Markovian, so that $\talpha_{nj}(t-s)$ is proportional to a delta function
Eq.~(\ref{MarkovBathCorrFuncPM}). 
Inserting Eq.~(\ref{MarkovBathCorrFuncPM}) into Eq.~(\ref{EvolEqForBSchlangeVonTWithBathCorrFct}) leads to
\begin{equation}
\label{EvolEqBSchlangeVonTMarkovian}
  \partial_t b_{nj}(t)=-i\Omega_{nj} b_{nj}(t)-\gamma_{nj} b_{nj}(t)-\frac{i}{\hbar}\tA_{nj}(t).
\end{equation}
Integration of Eq.~(\ref{EvolEqBSchlangeVonTMarkovian}) yields
\begin{equation}
\label{FinalExprBSchlangeVonT}
\begin{split}
   b_{nj}(t)=&e^{-i\Omega_{nj}t-\gamma_{nj}t} b_{nj}(0)\\
&-\frac{i}{\hbar}\int_0^t ds\ e^{-i\Omega_{nj}(t-s)-\gamma_{nj}(t-s)}\tA_{nj}(s).
\end{split}
\end{equation}
We now insert $ b_{nj}(t)$ Eq.~(\ref{FinalExprBSchlangeVonT}) into the correlation function
\begin{equation}
  \langle b_{nj}(t) b_{nj}^{\dagger}(s)\rangle_{T=0}=\bra{\hNull} b_{nj}(t) b_{nj}^{\dagger}(s)\ket{\hNull}
\end{equation}  
and obtain (by separately considering the cases $t<s$ and $t>s$)
\begin{equation}
\label{FinalCorrFctBSchlange}
  \langle b_{nj}(t) b_{nj}^{\dagger}(s)\rangle_{T=0}=e^{-i\Omega_{nj}(t-s)-\gamma_{nj}|t-s|}.
\end{equation}
Finally, Eq.~(\ref{FinalCorrFctBSchlange}) together with the definition of $B_n(t)$ 
from Eq.~(\ref{DefBfromBSchlange}) and Eq.~(\ref{bathCorrFuncTZeroPMElectr}) yields the result
\begin{equation}
\label{FinalCorrFctB}
  \beta_n(t-s)=\sum_j\Gamma_{nj}e^{-i\Omega_{nj}(t-s)-\gamma_{nj}|t-s|},
\end{equation} 
which is equal to the special bath correlation function $\alpha_n(t-s)$ 
given by Eq.~(\ref{BathCorrFuncForLorentzSpecDens}) for the NMQSD approach.

Thus, we have shown that the absorption spectrum we obtain from the PM method of 
section~\ref{sec:Exact_solv_model} is equal to the spectrum we get by using the NMQSD 
approach according to section~\ref{sec:qsd_approach}.

\begin{acknowledgments}
We thank John S.\ Briggs, Wolfgang Wolff, Alexander Croy, and Sebastian W\"uster for many helpful discussions.
\end{acknowledgments}


\end{document}